\documentclass[%
 amsmath,amssymb,
 aps,
]{revtex4-2}

\usepackage{graphicx}% Include figure files
\usepackage{bm}% bold math
\usepackage{natbib}
\usepackage{hyperref}
\hypersetup{
	colorlinks = true,
	urlcolor   = blue,
	citecolor  = blue,
}

\newcommand\dongdong[1]{\color{black}#1\color{black}}

\newcommand{\Ubz}{{U_b}_0}
\newcommand{\Ubo}{{U_b}_1}
\newcommand{\Ubt}{{U_b}_2}
\newcommand{\Ubth}{{U_b}_3}

\begin{document}

\preprint{APS/123-QED}

\title{Flow instability in Stokes layer of Carreau fluids}% Force line breaks with \\

\author{Mengqi Zhang}
\email{mpezmq@nus.edu.sg}
\altaffiliation[Currently at ]{Department of Mathematics, City University of Hong Kong, 83 Tat Chee Avenue, Kowloon Tong, Hong Kong.}%Lines break automatically or can be forced with \\
%\author{Dongdong Wan}%
%\email{ddwan@u.nus.edu}%\altaffiliation[Also at ]{State Key Laboratory for Strength and Vibration of Mechanical Structures and Department of Engineering Mechanics in School of Aerospace Engineering, Xi’an Jiaotong University, Xi’an 710049, PR China.}
\affiliation{
Department of Mechanical Engineering, National University of Singapore, 9 Engineering Drive 1, 117575 Singapore
}%

\author{\color{black} Dongdong Wan}%
\email{ddwan@xjtu.edu.cn}%\altaffiliation[Also at ]{}
\altaffiliation[\color{black}Previously at ]{Department of Mechanical Engineering, National University of Singapore, 9 Engineering Drive 1, 117575 Singapore.}
\affiliation{\color{black}State Key Laboratory for Strength and Vibration of Mechanical Structures, Department of Engineering Mechanics, Xi’an Jiaotong University, Xi’an, Shaanxi 710049, PR China}%

%\affiliation{%

\author{Huanshu Tan}%
\email{tanhs@sustech.edu.cn}
\affiliation{%
Multicomponent Fluids Group, Center for Complex Flows and Soft Matter Research, Department of Mechanics and Aerospace Engineering, Southern University of Science and Technology, Shenzhen, Guangdong 518055, PR China
}%

\date{\today}% It is always \today, today,

\begin{abstract}
This study investigates the influence of shear-thinning on the instability of a prototype time-periodic flow, the Stokes layer, in Carreau fluids. The time-dependent base flow was solved \dongdong{ using a numerical method and a binomial expansion method}. The expansion is conducted in terms of the nondimensional characteristic time ($\Lambda$), which quantifies the fluid's response time in viscosity to changes in shear rate. The expansion method shows good agreement with the numerical solution, provided that $\Lambda$ remains small. To understand the effect of shear-thinning on time-periodic flow instability, a Floquet analysis was conducted to examine two key parameters of the Carreau model, i.e., $\Lambda$ and the power-law exponent $n$. Our results show that decreasing $n$, which signifies stronger shear-thinning behavior, has a monotonic stabilizing effect on the flow within the range of investigated $n$. In contrast, increasing $\Lambda$ has a non-monotonic effect on the flow instability, which can be observed in both the weakly and strongly shear-thinning regimes. To clarify the instability mechanism, we perform an energy analysis showing that instability arises when the perturbation field is in phase with the oscillatory base flow, enabling efficient energy extraction from the time-dependent shear. A phase mismatch suppresses this transfer and stabilises the flow. This mechanism parallels the classical energy-production process in steady shear flows, where streamwise and wall-normal velocity perturbations exhibit a characteristic phase difference. Crucially, it is identified here for the first time in a time-periodic shear flow.
%\begin{description}
%\end{description}
\end{abstract}

\maketitle

%\tableofcontents

\section{Introduction}
The Stokes layer describes the time-periodic flow of a thin fluid layer over a flat plate oscillating in its own plane \cite{Stokes1851Effect}. This flow serves as a prototypical model for investigating more complex time-periodic flows, which appear in both industrial and natural processes. Examples include flows in piston-driven internal combustion engines \cite{Oppenheim2004Combustion}, pulsatile flows in pipelines driven by reciprocating pumps \cite{Ccarpinlioglu2001Critical}, and hemodynamic flow in arteries and veins \cite{Ku1997Blood}. Extensive research has been devoted to examining the stability, flow transition, and the onset of turbulence in time-periodic flows, primarily focusing on Newtonian fluids \cite{Davis1976}. However, many fluids in real-world applications are non-Newtonian, with shear-thinning fluids being particularly prevalent. Such fluids are commonly encountered in industrial applications (e.g., lubricants and polymer solutions), food products (e.g., ketchup and honey), and biological fluids (e.g., mucus and blood). In this study, we aim to investigate the effects of shear-thinning behavior on the stability of time-periodic systems and elucidate the underlying instability mechanism. In the following, we will first review the Stokes layer of Newtonian fluids and then discuss the shear-thinning effect in general.

\subsection{Stability/instability of the Stokes layer flow of Newtonian fluids}
In the literature, both semi-infinite and finite Stokes layer flows have been investigated \cite{Davis1976}. Early theoretical studies on the Stokes layer include Floquet analyzes in the semi-infinite configuration \cite{Hall1978Linear} and in the finite channel \cite{VonKerczek1974Linear}. These studies found no instability within a limited parameter space, mainly due to the restricted computational resources at that time. It is Blennerhassett \& Bassom \cite{Blennerhassett2002} who first found the onset of flow instability at a Reynolds number $Re$ of around 708 using the Floquet theory. The $Re$ is defined based on the Stokes layer thickness and the maximum wall oscillating velocity. They discovered both oscillatory and stationary unstable modes beyond the instability threshold, leading to a complex neutral stability curve. Hall showed analytically that the Floquet-type instability would eventually cease to exist at much higher Reynolds numbers \cite{Hall2003Instability}, a result also suggested by Cowley \cite{Cowley1987High}. Blennerhassett \& Bassom \cite{Blennerhassett2006} extended their study to finite Stokes layers in a channel and found that reducing the channel width first destabilized the flow, before stabilizing it. These linear instabilities were later confirmed through an instantaneous stability analysis using time-frozen base flows \cite{Luo2010Linear}; see also the momentary stability analysis in Ref. \cite{Blondeaux2021}. More recently, a transient growth analysis \cite{Biau2016} and a Finite-Time Lyapunov Exponent (FTLE) analysis \cite{Zhang2025} to the Stokes layer reveal significant energy amplification within the first oscillation cycle. The intracyclic instability identified via FTLE shows favorable agreement with the axial turbulence intensity in experiments \cite{Zhang2025}. Numerical simulations have also been conducted to study the flow stability/instability in Stokes layers, including wall roughness effect \cite{Vittori1998,Blondeaux1994,Costamagna2003,Luo2010Linear}, wavepacket evolution \cite{Thomas2014} and temporal evolution of 2D/3D disturbance in the Stokes layer \cite{Thomas2010}.

Despite the theoretical and numerical efforts, experiments predicted the instability onset at a lower transition Reynolds number compared to the Floquet analysis, lying in the range from 140 to 300, according to Refs. \cite{Hino1976Experiments,Hino1983,Jensen1989,Akhavan1991Investigation2}. Key hallmarks of transitional and turbulent flow have been documented in prior studies. For instance, Merkli \& Thomann \cite{Merkli1975Transition} reported intermittent turbulent bursts and subsequent relaminarization over a full oscillatory cycle. In experimenting with the fully turbulent oscillatory flow, its occurrence during the decelerating phase, followed by relaminarization in the accelerating phase, was also observed \cite{Hino1976Experiments}. These findings were later corroborated by \cite{Winter1984Turbulence}, who demonstrated that sustained turbulence requires a non-zero mean flow. Further insights were provided by \cite{Akhavan1991Investigation2}, who experimentally established that turbulence initiates as localized bursts in the near-wall region toward the end of the accelerating phase and persists throughout deceleration. Additionally, the same authors complemented their experimental work with direct numerical simulations of oscillatory channel flow \cite{Akhavan1991Investigation}, confirming that the transitional Reynolds number and turbulent structures align with their pipe flow observations.

It is noted that the experiments studying the oscillatory flow instability were consistently conducted in an oscillatory pressure-gradient driven flow, rather than the oscillating walls. Nevertheless, the two flows are mathematically equivalent to each other with a simple change of frame of reference. Thus, the instability in a pulsatile channel flow has also been studied theoretically in the literature as a prototypical case of oscillatory flow dynamics. The earliest work in this setup was conducted by \cite{Grosch1968Stability}, who showed that weak modulations of the driven pressure gradient were stabilizing, while strong modulations could drastically destabilize the flow, demonstrating a monotonic effect. In contrast, Von Kerczek \cite{VonKerczek1982Instability} uncovered a non-monotonic relationship between pulsation frequency and flow stability. More recently, Pier \& Schmid \cite{Pier2017Linear} systematically investigated the linear and finite-amplitude waves in the flow, revealing different nonlinear dynamics strongly depending on the pulsation amplitude and frequency. The same authors also calculated the optimal transient growth of the perturbation energy in the pulsating flow \cite{Pier2021}.

\subsection{Shear-thinning effects}
The previously reviewed works focus exclusively on Newtonian fluids. However, shear-thinning is a key rheological property of many natural and industrial fluids. A prominent example is blood, which exhibits pronounced shear-thinning behavior despite plasma alone being Newtonian. Preserving the normal shear-thinning characteristics of blood is essential for maintaining health, as deviations can lead to adverse physiological outcomes. For example, Alexy et al. \cite{Alexy2022Physical} demonstrated that the deformability of red blood cells plays a central role in modulating the shear-thinning response; alterations in this property are associated with a range of complex clinical conditions.

Despite its significance, the effect of shear-thinning on the stabilities of various flows has received less attention than expected, likely due to the complexity of the underlying physical mechanisms. Most relevant works to our work is the shear-thinning channel flow without oscillation. Motivated by its effect on drag reduction, Nouar et al. \cite{Nouar2007Delaying} explored the linear instability of the shear-thinning in Poiseuille flows of Carreau fluids. They found that selecting an appropriate characteristic viscosity scale in the definition of $Re$ was important for explaining the shear-thinning effect. Particularly, when an average viscosity across the channel was employed, the flow was first stabilized and then destabilized with increasing the shear-thinning effect. In another way when a tangent viscosity was used, the effect became consistently stabilizing. Based on a scaling analysis of the critical layer and boundary layer, they justified the adoption of the tangent viscosity as a more appropriate viscosity scale. Accordingly, they concluded that viscosity stratification could be an effective strategy to delay the transition to turbulence in channel flow, in contrast to \cite{Chikkadi2005}. Later, their work was extended to the weakly nonlinear regime \cite{Chekila2011}, where they showed that as the shear-thinning effect strengthened, the subcritical behavior of the flow system was further enhanced. The stabilizing effect of shear-thinning has also been suggested by Griffiths et al. \cite{Griffiths2014} in a rotating boundary layer flow, though they adopted the power-law fluid model instead. Refs. \cite{Arosemena2021,Arosemena2021a} further reported the shear-thinning effect on the turbulent channel flow. In the current study of the Stokes layer in a confined periodic channel using the Carreau model, since the base flow is time-dependent and so is the corresponding viscosity, our results will show more complicated effects of shear-thinning.

\subsection{The aim of the current study}
The literature review above highlights the complex stability and instability characteristics of Stokes layer flows in Newtonian fluids and the shear-thinning effect on the classic channel flow. However, despite the importance of shear-thinning fluids, their effects on time-periodic flows remain largely unexplored. The current study aims to address this gap by performing a Floquet analysis of Stokes layer flows in a channel with the Carreau rheological fluids. We seek to characterize the shear-thinning induced flow instability in the relevant parameter space and explain the physical mechanism behind the found instability. We do note that there are scattered studies on the Stokes layer of other complex fluids, such as Stokes layer of a viscoplastic fluid \cite{Balmforth2009a}, an elasto-viscoplastic fluid \cite{Hewitt2024} and a viscoelastic fluid \cite{Ortin2020}. Nevertheless, all of them focused on one-dimensional steady flows, instead of conducting Floquet analysis of the perturbation field. Thus, our work complements the existing literature on the common shear-thinning fluids by providing additional insights into their behavior.

The paper is organized as follows: Section \ref{problemformulation} presents the physical problem and mathematical formulations, including governing equations, control parameters and eigenvalue problem in the Floquet analysis. Two different methods employed for the calculation of the laminar base flow are described in Section \ref{baseflowderivation}. The results are provided and discussed in Section \ref{results}, followed by a summary and conclusion in Section \ref{Conclusions}.

\section{Problem formulation}\label{problemformulation}

\begin{figure}
	\centering
	\includegraphics[width=0.65\textwidth,trim= 0 0 0 0,clip]{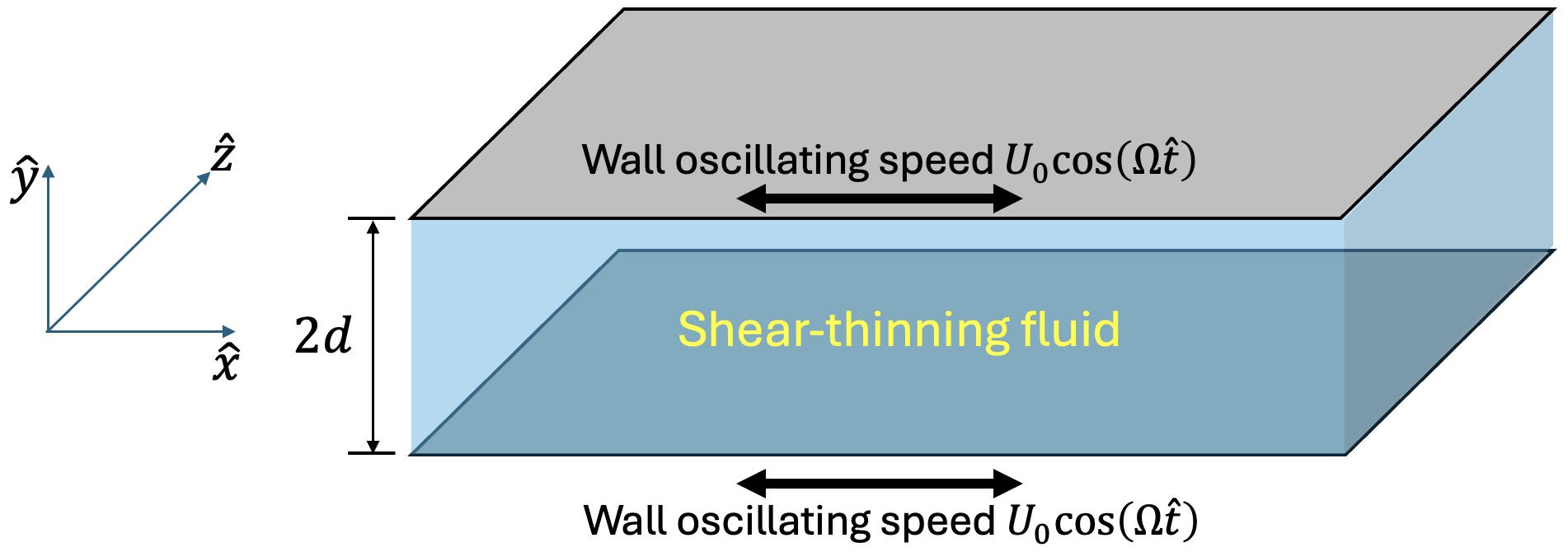}
	\caption{\color{black}Schematic of the Stokes layer flow of shear-thinning fluids modeled by the Carreau model in an oscillating channel. The channel walls move at a harmonic velocity $U_0\cos(\Omega \hat t)$ in their own planes along the $\hat x$ direction. }
	\label{Fig:Stokes_layer}
\end{figure}

{\color{black}In our Stokes layer as shown in FIG. \ref{Fig:Stokes_layer}, the walls move at a harmonic velocity $U_0\cos\Omega \hat t$ in their own planes, where $U_0$ is the maximum oscillating velocity and $\Omega$ is the dimensional oscillating frequency}. In experiments, it is more feasible to use piston-driven oscillating pressure gradient to drive Stokes layer flows than oscillating the walls of flow ducts. Note that these two problems are mathematically equivalent which can be proved via a simple coordinate transformation of the governing equations in the fixed frame of reference to a non-inertial frame of reference moving with the oscillating wall; also see section 4 of Ref. \cite{Blennerhassett2006}. In this theoretical study, we follow the conventional wall oscillation to study the Stokes layer flow dynamics.

The two walls are separated by $2d$ with the Cartesian coordinate located at the channel center. For sufficiently distanced walls, this flow mimics the dynamics of the classic semi-infinite Stokes layer. The streamwise, wall-normal and spanwise directions are denoted by $(\hat x,\hat y,\hat z)$, corresponding to the velocity components $\hat u,\hat v,\hat w$, respectively. The streamwise and spanwise wavenumbers are represented by $\alpha$ and $\gamma$ respectively. The mark hat indicates dimensional variables.

This work considers a shear-thinning Stokes layer. The shear-thinning behavior of the fluid is modeled by the Carreau model, accounting for the viscosity variation.
We would like to emphasise that the Carreau model has been extensively and consistently employed in the literature. Recent studies using this model include \cite{Chikkadi2005,Nouar2007Delaying,Chekila2011,Arosemena2021,Arosemena2021a,Plaut2017,Boyko2021,Alam2021,Henry2022,Singh2023,Ashkenazi2025,Milocco2025}, which numerically and experimentally investigate linear instability, weakly nonlinear dynamics, and turbulence in Carreau fluids. Notably, Singh \textit{et al.}~\cite{Singh2023} applied the Carreau model to a time-dependent pulsatile flow. These examples illustrate that the Carreau model is widely regarded as an appropriate and reliable framework for capturing shear-thinning behaviour. In light of this, we believe that its use in the present problem is both natural and well justified. We acknowledge, however, that the Carreau model does not incorporate elastic effects, which could be explored in future work.

\subsection{Governing equations}\label{governing_equations}
{\color{black}This incompressible fluid flow is governed by the following dimensional continuity equation and Navier-Stokes (NS) equation with the Carreau model
\begin{subequations}\label{NSEprimitive_dimensional}
\begin{equation}
\frac{\partial \hat{\bm u }}{\partial \hat t} + (\hat{\bm u}\cdot \hat\nabla )\hat{\bm u} = -\frac{1}{\rho}\hat\nabla \hat p + \hat\nabla\cdot\hat{\bm \tau}, \ \quad \quad \hat{\bm \tau}=\hat\nu \hat{\dot{\boldsymbol \gamma}}, \ \quad \quad \hat\nabla\cdot\hat{\bm u}=0,\label{NSE}
\end{equation}
\begin{equation}
\text{coupled with the Carreau model }  \ \ \ \ \  \hat\nu = \hat \nu_\infty+(\hat\nu_0-\hat\nu_\infty) \big[  1+ (\lambda \hat{\dot \gamma})^2 \big]^\frac{n-1}{2},\label{baseCarreau}
\end{equation}
\end{subequations}
where $\hat{\bm u }$ is the velocity vector, $\hat{p }$ is pressure, $\hat{\bm \tau }$ is stress tensor (divided by the density $\rho$) with $\hat{\dot{\boldsymbol \gamma}}=\hat\nabla \hat{\bm u}+(\hat\nabla\hat{\bm u})^T$. Regarding the Carreau model, $\hat\nu$, $\hat\nu_\infty$ and $\hat\nu_0$ are the kinematic viscosity, infinite-shear-rate viscosity and zero-shear-rate viscosity, respectively. The characteristic time scale $\lambda$ measures the response time of the viscosity to the change in shear rate $\hat{\dot \gamma} = \sqrt{\frac{1}{2}\hat{\dot{\boldsymbol \gamma}} :\hat{\dot{\boldsymbol \gamma}} }$. The power-law index $n$ is a model fitting parameter to be discussed below. }

{\color{black} For non-dimensionalization of shear-thinning fluid flow systems, one typical way is to select the channel half height $d$ (or tube radius for flows in tubes), but it is mainly suitable for steady or pulsatile flows with a steady flow component; see for example Refs. \cite{Baez-Amador2024,Chun2024Flow}. The present study focuses on oscillating channel flows of shear-thinning fluids with no steady flow component and the flow is mainly restricted within the thin Stokes layers close to the walls.} Therefore, we follow Refs. \cite{Blennerhassett2002,Blennerhassett2006} instead to nondimensionalize the flow system using the length scale $\sqrt{2\nu_0/\Omega}$, which is the Stokes layer thickness. In addition, we select  the velocity scale $U_0$, the time scale $1/\Omega$ and the pressure scale $\rho U_0^2$, where $\nu_0$ is the zero-shear kinematic viscosity coefficient and $\rho$ is the density. Subsequently, the nondimensionalized profile of the wall oscillation becomes $\cos t$ and the nondimensional incompressible Navier-Stokes (NS) equations read
\begin{subequations}\label{NSEprimitive}
\begin{equation}
\frac{\partial \tilde{\bm u }}{\partial t} + Re (\tilde{\bm u}\cdot \nabla )\tilde{\bm u} = -Re\nabla \tilde p + \frac{1}{2}\nabla\cdot\tilde{\bm \tau}, \ \quad \quad \tilde{\bm \tau}=\tilde\nu \tilde{\dot{\boldsymbol \gamma}}, \ \quad \quad \nabla\cdot\tilde{\bm u}=0,\label{NSE}
\end{equation}
\begin{equation}
\text{coupled with the Carreau model }  \ \ \ \ \  \tilde\nu = \nu_\infty+(1-\nu_\infty) \big[  1+ (Re\lambda \Omega \tilde{\dot \gamma})^2 \big]^\frac{n-1}{2},\label{baseCarreau}
\end{equation}
\end{subequations}
where the infinite-shear-rate viscosity $\nu_\infty$, {\color{black} describing the shear-thinning limit of the fluid}, has been nondimensionalized by $\nu_0$. Since the infinite-shear-rate viscosity is generally much smaller than the zero-shear-rate viscosity, \dongdong{the nondimensional $\nu_\infty$ is set to be zero throughout this study}, following Ref. \cite{Nouar2007Delaying}. In the above equations, the Reynolds number is defined as $Re=\frac{U_0}{\sqrt{2\nu_0\Omega}} $ based on the Stokes layer thickness.
\dongdong{ It characterizes the importance of fluid inertia due to the oscillating walls relative to the viscous effect. Typically, the laminar Stokes layer flow at high $Re$ is prone to instability triggered by infinitesimal disturbances for Newtonian fluids \cite{Blennerhassett2006}. } 
The Reynolds number can also be written as $Re=\frac{1/\Omega}{\sqrt{2\nu_0/\Omega}/U_0}$, which can be interpreted as the ratio of the time scale for wall oscillation $1/\Omega$ to the time scale for the 'penetration' effect $\sqrt{2\nu_0/\Omega}/U_0$. For the Carreau model, we will denote $Re\lambda\Omega$ in Eq. (\ref{baseCarreau}) collectively as
\begin{align}\label{eq:Lambda_def}
\Lambda=Re\lambda \Omega =  \frac{\lambda}{\sqrt{2\nu_0/\Omega}/U_0}, 
\end{align}
which can be interpreted as the ratio of the viscosity relaxation time to the wall-oscillation penetration time. When this parameter is small, the behavior of the fluids resembles its Newtonian counterpart in the Stokes layer; whereas it is large, the non-Newtonian effect becomes stronger.
In the Carreau model, the power-law index $n$ also characterizes the degree of the non-Newtonian effect; the larger the deviation from 1, the stronger the non-Newtonian effect. Typically, shear-thinning fluids (e.g., polymer solutions) correspond to $0.2<n<1$ \cite{Bird1987} {\color{black} and small $n$ indicates a narrow transitional range from $\nu_0$ to $\nu_\infty$}.  In contrast, $n>1$ manifests as shear-thickening effect. Shear-thickening fluids (like dense colloidal suspensions, cornstarch-water mixtures) often show abrupt or discontinuous viscosity increases at certain shear rates \cite{Seto2013,Ness2022}. \dongdong{The Carreau model is smooth and monotonic, and cannot capture sharp jumps. Therefore, we exclusively focus on the Carreau model with $n<1$ for shear-thinning fluids in our work.}

For the Newtonian fluids, driven by the periodic wall motion mentioned above, the equations admit a classic analytical solution of a time-periodic flow $\bm U_b$ in the $x$ direction
\begin{align}\label{baseUNew}
{\bm U_b}=(\Ubz(y,t),0,0) = (U_0(y) e^{i t} + U_0^*(y)e^{-i t} , 0 ,0 ), \ \ \  \text{  with   }\ \ \ U_0(y) =\frac{  \cosh{\sqrt{2i}y}   }{  2\cosh\sqrt{2i}h  },
\end{align}
where $i$ is the imaginary unit and the superscript $^*$ marks the complex conjugate. $h = \frac{d}{\sqrt{2\nu_0/\Omega}}$ denotes the nondimensional channel half-height, {\color{black} similar to Womersley number which is more widely used for pulsatile flows with a stead flow component, especially in biofluid mechanics. Large $h$ corresponds to channels with widely separated Stokes layers. Effect of $h$ is generally non-monotonic and instability is mostly easily triggered at intermediate values of $h$ for Newtonian fluids \cite{Blennerhassett2006}.} 

For the Carreau fluids, due to its intrinsic nonlinearity in Eq. \eqref{baseCarreau}, a similar analytical solution is currently not available. In this work, we will resort to two methods; one is a numerical method to solve for all the harmonics involved in the solution and the other is an expansion method to resolve the low-order harmonics, to be explained in Sec. \ref{baseflowderivation}. 

\subsection{Linearized equations}
Assuming that the time-periodic laminar base flow has been obtained, the linear analysis follows by performing the Reynolds decomposition of the flow variables, i.e. \dongdong{$\tilde{\bm u}({\bm x},t)={\bm U_b}({\bm x}, t) + {\bm u}({\bm x},t), \tilde p({\bm x},t)=P_b({\bm x},t) + p({\bm x},t), \tilde\nu({\bm x}, t) = \nu_b({\bm x}, t)  + \nu({\bm x}, t) , \tilde{\dot{\boldsymbol \gamma}}({\bm x}, t) = {\dot{\boldsymbol \gamma}}_b({\bm x}, t)  + {\dot{\boldsymbol \gamma}}({\bm x}, t) $. The base flow variables are marked with the subscript ``b''. They also satisfy Eq. \eqref{NSEprimitive} and the governing equation can be written as
\begin{equation}\label{Eqbaseflow}
\frac{\partial {\bm U_b }}{\partial t} + Re ({\bm U_b}\cdot \nabla ){\bm U_b} = -Re\nabla  P + \frac{1}{2}\nabla\cdot{\bm \tau_b}, \   {\bm \tau_b}=\nu_b {\dot{\boldsymbol \gamma_b}}, \  \nu_b =  \big(  1+ \frac{\Lambda^2}{2} \dot {\boldsymbol\gamma}_b: \dot{\boldsymbol\gamma}_b \big)^\frac{n-1}{2}, \  {\dot{\boldsymbol \gamma}_b}=\nabla {\bm U_b}+(\nabla{\bm U_b})^T, \ \nabla\cdot{\bm U_b}=0.
\end{equation}
With the introduction of the classic analytical solution form ${\bm U_b}=(U_b(y,t),0,0)$ and $P_b=0$, we have $\dot{\boldsymbol \gamma}_b  = [0\ \ U_b'\ \ 0 ; U_b'\ \ 0\ \ 0; 0\ \ 0\ \ 0 ]$ and then Eq. \eqref{Eqbaseflow} further simplifies to
\begin{align}
 \frac{\partial U_b}{\partial t}  &=  \frac{1}{2} \frac{\partial }{\partial y}( \nu_b \frac{\partial U_b}{\partial y}).  \label{baseflowEq}
\end{align}
The lower-case symbols without tilde represent the infinitesimal perturbation of the flow quantities. By substituting the Reynolds decomposition, expanding the terms, neglecting nonlinear terms and subtracting the base-flow terms, we arrive at
\begin{align}\label{linearizedeq}
\frac{\partial  {\bm u}}{\partial t} + Re[ {\bm U_b} \cdot \nabla  {\bm u} +  {\bm u} \cdot \nabla {\bm U_b}  ] = -Re\nabla p + \frac{1}{2}\nabla \cdot  {\bm \tau}, \quad\quad \nabla \cdot {\bm u}=0,
\end{align}
where the no-slip boundary conditions are applied to the velocity components. The perturbed stress tensor in the above equation reads
\begin{subequations}
\begin{equation}
\bm\tau=\nu_b  \dot{\boldsymbol \gamma} + \nu \dot{\boldsymbol \gamma}_b, \quad \dot{\boldsymbol \gamma}  = \nabla {\bm u}+(\nabla{\bm u})^T,
\end{equation}
\begin{equation}
\nu  = \frac{\partial \nu_b}{\partial \dot{\boldsymbol\gamma}_b}: \dot{\boldsymbol\gamma} = \frac{n-1}{2}\big[  1+  \frac{\Lambda^2}{2} \dot {\boldsymbol\gamma}_b: \dot{\boldsymbol\gamma}_b  \big]^\frac{n-3}{2} \Lambda^2\dot{\boldsymbol\gamma}_b:\dot{\boldsymbol\gamma},\label{eq25}
\end{equation}
\end{subequations}}
where $\dot{\boldsymbol\gamma}_b:\dot{\boldsymbol\gamma}=2U_b'(\frac{\partial u }{\partial y}+\frac{\partial v}{\partial x})$ and the prime denotes the $y$-derivative (so are the symbols $D,D^2$ in the following text). Following Ref. \cite{Nouar2009}, for the unidirectional flow, one can obtain 
\begin{subequations}\label{nubnubt_def}
\begin{equation}
\tau_{ij}=\nu_b  \dot{\boldsymbol \gamma}, \hspace{3.8cm} \text{when $ij\ne xy,yx$},
\end{equation}
\begin{equation}
\tau_{ij}={\nu_b}_t\dot{\boldsymbol \gamma}, \ \ \ \ {\nu_b}_t=\nu_b+\frac{\partial \nu_b}{\partial \dot{\boldsymbol\gamma}_b}: \dot{\boldsymbol\gamma}_b, \ \ \ \text{when $ij= xy,yx$,}
\end{equation}
\end{subequations}
where ${\nu_b}_t$ is called the base tangent viscosity coefficient. The profiles of ${\nu_b}$ and ${\nu_b}_t$ in the Poiseuille flow of Carreau fluids have been reported in \cite{Nouar2007Delaying}. Our calculation of the base flow velocity $U_b$ and viscosity $\nu_b,{\nu_b}_t$ in that flow can be found in Appendix \ref{validationappendix} and compared favourably to theirs. 

The linearized equations (\ref{linearizedeq}) can be reformulated in terms of the vertical velocity $v$ and the vertical vorticity $\eta= \frac{\partial u}{\partial z} - \frac{\partial w}{\partial x}$ by utilising the continuity equation to eliminate the pressure gradient term. The $v-\eta$ formulation in the Stokes layer of Carreau fluids reads 
\begin{subequations}
\begin{equation}
\frac{\partial \nabla^2 v}{\partial t} = -  ReU_b \frac{\partial \nabla^2 v}{\partial x}  + ReU''_b\frac{\partial v}{\partial x}  +     \frac{1}{2}\Big[ \frac{\partial^3\tau_{2i}}{\partial x_j^2\partial x_i} -  \frac{\partial^3 \tau_{ij}}{\partial y\partial x_i\partial x_j} \Big],
\end{equation}
\begin{equation}\label{eq_eta}
\frac{\partial  \eta}{\partial t} = - ReU_b \frac{\partial  \eta}{\partial x} - ReU'_b \frac{\partial v}{\partial z}     + \frac{1}{2}\frac{\partial}{\partial x_j}\Big[ \frac{\partial \tau_{1j}}{\partial z} - \frac{\partial \tau_{3j}}{\partial x} \Big].
\end{equation}
\end{subequations}
\dongdong{In the present study, we focus on 2-D problems and thus ignore equation (\ref{eq_eta}) and all the variations with respect to $z$ below. After substituting the expressions for the perturbed stress tensor $\bm\tau$ and assuming the wavelike assumption $v(x,y,t)={\check v}(y,t) e^{i\alpha x}$, where $\alpha$ stands for real-valued streamwise wavenumber, we arrive at
\begin{align}\label{eq2d8}
{i} \frac{\partial \nabla^2 \check v}{\partial t} = \alpha Re U_b \nabla^2 \check v & - \alpha Re U''_b\check v  + \frac{i}{2}  \Big[  ( D^2 + \alpha^2) \Big( ({\nu_b}_t-\nu_b) (D^2+\alpha^2)\check v \Big)     \nonumber \\
&+   \nu_b \Delta^2\check v   + 2 D\nu_b D^3 \check v+   D^2  \nu_b D^2\check v     -2\alpha^2 D \nu_b D\check v  +   \alpha^2  D^2  \nu_b\check v  \Big].
\end{align}
The same equation (except the difference due to the nondimensionalization) has been previously obtained by \cite{Nouar2009} in the context of the shear-thinning channel flow. Since the coefficients in the linearized problem is time-periodic, we seek for a solution based on the Floquet theory \cite{Coddington1955}. 
The Floquet-Fourier-Hill (FFH) method dictates a solution form for the shape function as $ {\check v}(y,t)= \displaystyle \Big[\sum_{m=-\infty}^\infty {v}^{(m)}(y) e^{imt} \Big] e^{-i\omega t}$ based on a Fourier transform. Alternatively, we can write collectively the flow variables as
\begin{align}\label{solutionform}
{v}(x,y,t) = \Big[\sum_{m=-\infty}^\infty {v}^{(m)}(y) e^{imt}\Big] e^{i\alpha x  -i\omega t}+ c.c.,
\end{align}
where $c.c.$ means complex conjugate of all the preceding terms. {\color{black} The complex frequency $\omega=\omega_r + i\omega_i$ encompasses the growth rate $\omega_i$ and frequency $\omega_r$ of the disturbance}. Physically, it signifies the amplification/decay of the disturbance over a period $T$. In studying the linear stability of Newtonian Stokes layers, this method has been adopted in Refs.  \cite{Blennerhassett2002,Blennerhassett2006} in semi-infinite space and in a channel, respectively.

By substituting Eq. \eqref{solutionform} into Eq. \eqref{eq2d8}, we can obtain an infinite set of equations for each Fourier mode ${v}^{(m)}$
\begin{align}\label{eq:v}
   \omega  v^{(m)} & = m  v^{(m)}  + \alpha Re\nabla^{-2} \sum_{j=-N_f}^{N_f}  \Big[ U_j   \nabla^2 v^{(m-j)} - U_j'' v^{(m-j)}  +     U_j^* \nabla^2 v^{(m+j)} -  {U_j^*}'' v^{(m+j)} \Big] \nonumber\\
&    +  \frac{i}{2}\nabla^{-2} \sum_{j=-N_f}^{N_f}  \Big[ ( D^2 + \alpha^2) \Big( ({{\nu_b}_t}_j - {\nu_b}_j) (D^2+\alpha^2) v^{(m-j)} + ({{\nu^*_b}_t}_j - {\nu^*_b}_j) (D^2+\alpha^2) v^{(m+j)} \Big)     \nonumber \\
&+   {\nu_b}_j \Delta^2 v^{(m-j)}  + {\nu^*_b}_j \Delta^2  v^{(m+j)}   + 2 D{\nu_b}_j D^3  v^{(m-j)}  + 2 D{\nu^*_b}_j D^3 v^{(m+j)}   \nonumber\\
& +   D^2  {\nu_b}_j D^2  v^{(m-j)}   +   D^2  {\nu^*_b}_j D^2 v^{(m+j)}  - 2\alpha^2 D {\nu_b}_j D v^{(m-j)}  - 2\alpha^2 D {\nu^*_b}_j D v^{(m+j)}   \nonumber  \\
&+   \alpha^2  D^2  {\nu_b}_j  v^{(m-j)} +   \alpha^2  D^2  {\nu^*_b}_j  v^{(m+j)}  \Big],
\end{align}
where $U_j$ is the $j-$th Fourier component of the time-periodic base flow and ${\nu_{b}}_j, {\nu_{bt}}_j$ should be interpreted similarly.

Numerically, we will truncate the infinite expansion $v$ and only retain the Fourier modes from $-N_f$ to $N_f$, resulting in the eigenvector ${\bm q}_T = [{v}^{(-N_f)}\  {v}^{(-N_f+1)} \ ... {v}^{(-1)} \ {v}^{(0)}\ {v}^{(1)} \  ... \  {v}^{(N_f-1)}\  {v}^{(N_f)} ]^T$. Correspondingly, the eigenvalue problem (\ref{eq:v}) can be written in a compact form of $\omega {\bm q}_T  = {\bm L}  {\bm q}_T $ where the linear operator ${\bm L}$ can be readily derived from (\ref{eq:v}). Sufficient large $N_f$ will be chosen to ensure the convergence of the Fourier expansion. Thomas et al. \cite{Thomas2011} found that $N_f$ should be approximately greater than $0.8\alpha Re$ in the Newtonian flow. Our computation shows that this remains valid for the Stokes layer flow of weakly shear-thinning Carreau fluids at large $n$ and small $\Lambda$, and may be also applicable for strongly shear-thinning fluids at small $n$ and large $\Lambda$; see Appendix \ref{convergence_appendix} for more data and discussions.}

In our analysis, the computational domain is discretized using a spectral collocation method \cite{Weideman2000,Trefethen2000} with $N_y$ interior grid points (excluding the boundary grid points) in the wall-normal direction. Thus, the dimension of $\bm q_T$ is $N_y(2N_f+1)$. The no-slip boundary conditions of the variables can be easily implemented by removing the first/last rows and first/last columns of the corresponding matrices in the collocation method. The clamped boundary conditions of $v$ have been implemented in the code scripts provided by \cite{Weideman2000}. 

\section{Base flow for the Stokes layer of shear-thinning fluids}\label{baseflowderivation}
As explained in the previous section, \dongdong{ the time-periodic laminar solution to the Eq. \eqref{baseflowEq} will be used as the base flow}. In contrast to the Newtonian case, a theoretical expression for this solution remains unavailable. Therefore, we resort to a numerical method and an expansion method, to be explained below.
\subsection{Numerical method}
In the numerical method, a spectral collocation method based on Chebyshev polynomials is adopted to discretize the flow in space. For the temporal evolution, we employ the trigonometric polynomials on equispaced ``time grid''. Subsequently, a circulant Toeplitz differentiation matrix and Chebyshev differentiation matrices can be constructed for the calculation of the temporal and spatial derivatives, respectively; see sections 1 and 6 in \cite{Trefethen2000}. The discretization results in a nonlinear equation system, which is then solved using an iterative Newton method, e.g., the \emph{fsolve} function in MATLAB. During the iterations, the time-periodic boundary conditions at the two walls are enforced by directly assigning proper values to the variables at the boundary nodes. Concerning the resolution, the typical number of the temporal nodes in our calculation is $N_t=2N_f$ (with $N_f=300$) and that of Chebyshev-Lobatto nodes is $N_y=101$. Our convergence study shows that this resolution is enough to achieve spectral accuracy.

With adequate resolution, the numerical method can be, in principle, applied to all values of \( \Lambda \) and \( n \). However, extending it to large \( \Lambda \) becomes increasingly challenging due to the strong nonlinearity, see the discussions on figure \ref{Fig:PSD_base_flow} below. On the other hand, we desire to derive some theoretical profile for the shear-thinning base flow using an expansion method as presented below, which can be useful and convenient for studies on the weakly shear-thinning (small $\Lambda$) flow regime, a regime that has received attention in the literature \cite{Khayat1996,Albaalbaki2008,Sun2017Flow}. By using this method, fully analytical solutions can be obtained for low order approximation up to the order of $\Lambda^2$; for the order of $\Lambda^4$ and above, semi-analytical solutions can be solved for via nonlinear iterations of only ordinary differential equations. The execution time of obtaining the analytical solution is much shorter than that of getting the numerical simulation. In addition, the obtained solutions can be compared to the fully numerical results as validation.

\subsection{Expansion method}
In the expansion method, we consider a small value of $\Lambda$, under which condition, the base state of viscosity $\nu_b$ in Eq. (\ref{baseCarreau}) can be simplified using a binomial approximation \dongdong{
\begin{align}
\nu_b =\big[  1+ \Lambda^2  (\dot {\boldsymbol\gamma}_b)^2  \big]^\frac{n-1}{2} =   1 + \frac{n-1}{2} \Lambda^2(\frac{\partial U_b}{\partial y})^2 + \frac{(n-1)(n-3)}{8} \Lambda^4(\frac{\partial U_b}{\partial y})^4 \label{nubEq} + \frac{(n-1)(n-3)(n-5)}{48} \Lambda^6(\frac{\partial U_b}{\partial y})^6 + \mathcal O(\Lambda^8).
\end{align}}
In this work, we have expanded the solution to $\Lambda^6$. We will evaluate the effect of the truncation order by comparing the expansion solution to the reference numerical solution.

Considering a small $\Lambda$, \dongdong{ one can assume the solution ansatz for solving Eq. \eqref{baseflowEq} to be } $U_b(y,t)= \Ubz(y,t) + \Lambda^2   \Ubo(y,t) + \Lambda^4   \Ubt(y,t)+ \Lambda^6   \Ubth(y,t) + O(\Lambda^8)$, consistent to the Eq. \eqref{nubEq}. We require that the boundary conditions for $\Ubz$ be the wall oscillating velocity $\cos t$ and that for all the other $U_{bj}$ (including $\Ubo,\Ubt$ and $\Ubth$) homogeneous. One may wonder why only even-order powers of \( \Lambda \) are retained in the expansion. The reason is that the odd-order terms, combined with the homogeneous boundary conditions, would result in a trivial solution, as can be easily proved.
By substituting the base-flow solution ansatz into Eq. (\ref{baseflowEq}) and utilising Eq. (\ref{nubEq}), one can obtain the following set of equations at various orders of $\Lambda$.

\subsubsection{At the order of $\Lambda^0$}
At the zeroth order, we have
\begin{align}\label{Lambda0Eq}
\frac{\partial  \Ubz }{\partial t}  &= \frac{1}{2}\frac{\partial }{\partial y}\Big( \frac{\partial  \Ubz }{\partial y} \Big), 
\end{align}
which is the same equation for the Newtonian fluid, as shown in Eq. (\ref{baseUNew}). 

\subsubsection{At the order of $\Lambda^2$}\label{sec:order_lambda2}
At this order, we have
\begin{align} \label{Lambda2Eq}
\frac{\partial  \Ubo}{\partial t}  = \frac{1}{2}\frac{\partial }{\partial y}\Big( \frac{\partial   \Ubo}{\partial y}+\frac{n-1}{2} (\frac{\partial  \Ubz }{\partial y})^3 \Big).
\end{align}
To analyze this equation, we notice that $(\frac{\partial  \Ubz }{\partial y})^3$ is the non-homogeneous part of the partial differential equation. Since $\Ubz=U_0(y) e^{i t} + U_0^*(y)e^{-i t}$, we anticipate that the cubic term in the above equation can generate harmonics $e^{it}$ and $e^{3it}$. Therefore, we assume $\Ubo(y,t) = U_{11}(y) e^{i t} + U_{13}(y) e^{3i t} + c.c.$ with $U_{11}, U_{13}$ to be determined. By substituting this ansatz into Eq. (\ref{Lambda2Eq}), we can obtain the equations for $U_{11}$ and $U_{13}$, respectively, 
\begin{subequations} 
\begin{eqnarray}
i U_{11} & =& \frac{1}{2}\Big( \frac{\partial^2    U_{11}}{\partial y^2}  -\frac{n-1}{2} 6i(\frac{\partial  U_0 }{\partial y})^2  U_0^*  + \frac{n-1}{2} 12i \frac{\partial  U_0 }{\partial y} \frac{\partial  U_0^* }{\partial y} U_0 \Big),
\end{eqnarray}
\begin{eqnarray}
3i U_{13} &= &\frac{1}{2}\Big(  \frac{\partial^2   U_{13}}{\partial y^2}+ \frac{n-1}{2} 6i(\frac{\partial  U_0 }{\partial y})^2 U_0 \Big).
\end{eqnarray}
\end{subequations} 
To solve these ordinary differential equations (ODEs), we start by determining the general solution to the homogeneous part of the equation. Once the homogeneous solution is obtained, we proceed to find the particular solution that satisfies the non-homogeneous part of the ODEs. The detailed solution procedure is shown in Appendix \ref{U11U13}. The solutions of $U_{11}$ and $U_{13}$ are given
\begin{subequations}\label{U11_U13}
\begin{eqnarray}
U_{11}(y) & = &  \frac{K}{10}(4+3i)\Big(\cosh(3+i)h\cosh{\sqrt{2i} y} - \cosh(3+i)y\cosh \sqrt{2i}h \Big)  \nonumber \\
 && -iK \Big( \cosh(1-i)h \cosh{\sqrt{2i} y} - \cosh(1-i)y\cosh \sqrt{2i}h \Big)  \\
 && - \frac{K}{10}(4-3i)\Big( \cosh(1+3i)h   \cosh{\sqrt{2i} y} -  \cosh(1+3i)y\cosh \sqrt{2i}h \Big),   \nonumber \\
K &= &\frac{n-1}{2} \frac{3}{16}(\frac{1 }{\cosh \sqrt{2i}h  })^3    \frac{1 }{\cosh \sqrt{2/i}h  }, \nonumber
\end{eqnarray}
\begin{eqnarray}
U_{13}(y) & =& \frac{n-1}{2}\frac{i}{8}  \Big[  \frac{ \cosh{\sqrt{6i} y}}{ \cosh{\sqrt{6i}h}}  -  \frac{\cosh^3(\sqrt{2i}y  ) }{  \cosh^3(\sqrt{2i}h)   }\Big ].
\end{eqnarray}
\end{subequations}

\subsubsection{At the order of $\Lambda^4$}
At this order, we obtain
\begin{align}\label{Lambda4Eq}
 \frac{\partial    \Ubt }{\partial t}  &= \frac{1}{2}\frac{\partial }{\partial y}\Big[ \frac{\partial \Ubt }{\partial y}+ \frac{n-1}{2} 3\big(\frac{\partial  \Ubz}{\partial y} \big)^2\big(\frac{\partial  \Ubo}{\partial y} \big)  + \frac{(n-1)(n-3)}{8} (\frac{\partial \Ubz }{\partial y})^5 \Big].
\end{align}
A procedure similar to that in the previous section \ref{sec:order_lambda2} can be followed by assuming $\Ubt(y,t) = U_{21} e^{it}+ U_{23} e^{3it}+ U_{25}e^{5it} + c.c.$, which leads to three ODEs as shown in Appendix \ref{app:order4}. Because manual derivation of the theoretical solution will be exceedingly difficult in this case, we will numerically construct the non-homogeneous terms on the right-hand-side of the equations and solve the resultant Helmholtz equations for $U_{21}$, $U_{23}$ and $U_{25}$.

%\subsubsection{At the order of $\Lambda^6$ }
\dongdong{
The analysis at the order of $\Lambda^6$ and the solution for $U_{b3}$ are described in Appendix \ref{app:order6}. 
With $\Ubz,\Ubo,\Ubt,U_{b3}$ being obtained theoretically and numerically, we can summarize the time-periodic base flow for the Stokes layer of Carreau fluids for small $\Lambda$ (involving only the first several harmonics) as 
\begin{align} \label{baseflowexpand}
U_b(y,t) &= \Ubz(y,t) + \Lambda^2  \Ubo(y,t)  + \Lambda^4   \Ubt(y,t)+ \Lambda^6   \Ubth(y,t) + ...
%\text{where} \ \ \  \Ubz(y,t) & = U_0(y) e^{i t} + c.c. \nonumber  \\
%\Ubo(y,t) & = U_{11}(y) e^{i t} + U_{13}(y) e^{3i t} +c.c.  \nonumber \\
%\Ubt(y,t) &= U_{21}(y) e^{it}+ U_{23}(y) e^{3it}+ U_{25}(y) e^{5it} + c.c. \nonumber \\
%U_{b3}(y,t) &= U_{31}(y) e^{it}+ U_{33}(y) e^{3it}+ U_{35}(y) e^{5it}+ U_{37}(y) e^{7it} + c.c.
\end{align}
In terms of the harmonics, the above base flow $U_b(y,t)$ can be equivalently written as
\begin{subequations} \label{twoharmonicbase}
\begin{eqnarray}
U_b(y,t) &=& [U_1(y) e^{it} + c.c.]+ [U_3(y) e^{3it}+ c.c.] + [U_5(y)e^{5it} + c.c.]+ [U_7(y)e^{7it}+ c.c.]  + ... \\
\text{with} \ \  U_1 &=& U_0 + \Lambda^2 U_{11} + \Lambda^4 U_{21} + \Lambda^6 U_{31}, 
U_3 = \Lambda^2 U_{13}+ \Lambda^4 U_{23} + \Lambda^6 U_{33}, 
 U_5 =  \Lambda^4 U_{25} + \Lambda^6 U_{35},
  U_7 =  \Lambda^6 U_{37}.
\end{eqnarray}
\end{subequations}}

Theoretically, one can retain more terms in the expansion, but this procedure will soon become manually intractable. We will compare the theoretical and numerical results for the base state in the limit of small $\Lambda$ and determine the range of $\Lambda$ in which the expansion solution is accurate compared to the reference numerical method. Then, the linear stability analysis of these two types of base flows will be conducted to quantify the difference between them.  

On the other hand, the viscosity coefficients $\nu_b$ and ${\nu_b}_t$ are also time-dependent. For these quantities, we will use the original equations \eqref{eq25} and \eqref{nubnubt_def}, respectively, with the numerical or expanded base flows.

\section{Results and Discussion} \label{results}
In the following, we will first present the base flow profiles and compare the two methods used to obtain them. Then the eigenvalues and eigenvectors in the Floquet analysis will be calculated and analyzed, followed by a discussion on the neutral curves. A parametric study will reveal the effects of various parameters on the flow instability. In the end, we will conduct an energy analysis to understand the instability mechanism in the Stokes layer of Carreau fluids. We fix $h=5$ in this work.

\subsection{Base flow profiles}
\begin{figure}
	\centering
	\includegraphics[width=0.99\textwidth,trim= 0 0 0 0,clip]{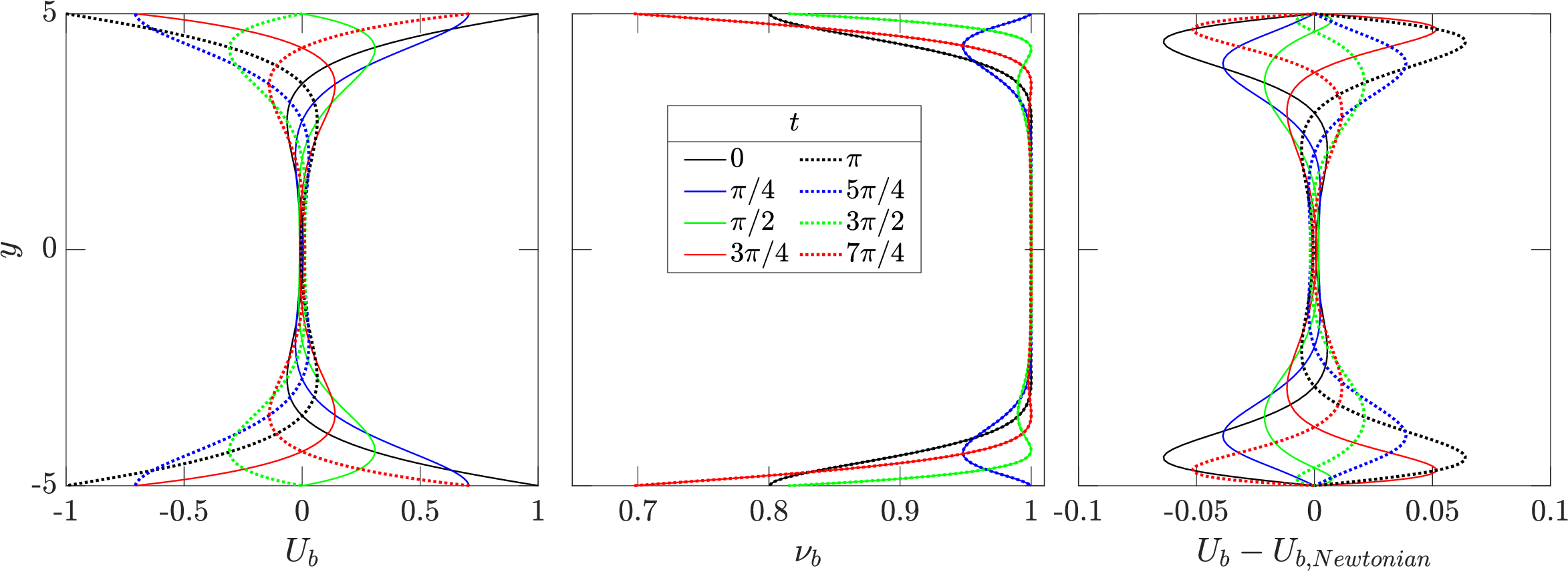} 
	\put(-508,185){$(a)$}\put(-323,185){$(b)$}\put(-158,185){$(c)$}
	\caption{Laminar base flow of the Stokes layer of Carreau fluids at $n=0.5$ and $\Lambda=1$. $(a)$ Profiles of the velocity $U_b$, $(b)$ viscosity $\nu_b$, $(c)$ the velocity difference between the shear-thinning and the Newtonian fluids $(U_b-U_{b,Newtonian})$. The lines represent the time stamps at eight different instants in a $2\pi$ period. }
	\label{Fig:base_flow_n05lambda1}
\end{figure}

Figure \ref{Fig:base_flow_n05lambda1} $(a,b)$ plot typical velocity profiles $U_b$ of the laminar base flow and the associated viscosity profiles $\nu_b$, respectively, at $n=0.5$ and $\Lambda=1$ as an example. Profiles at eight time instants $t$ are shown to demonstrate the evolution in a complete $2\pi$ period. Shear-thinning is evident from panel $(b)$ where close to the oscillating walls the viscosity coefficient $\nu_b$ deviates from one. Panel $(c)$ depicts the velocity difference $U_b-U_{b,Newtonian}$ between the Newtonian and shear-thinning fluids. The main difference lies in the near-wall region.

\begin{figure}
	\centering
	\includegraphics[width=0.99\textwidth,trim= 0 0 0 0,clip]{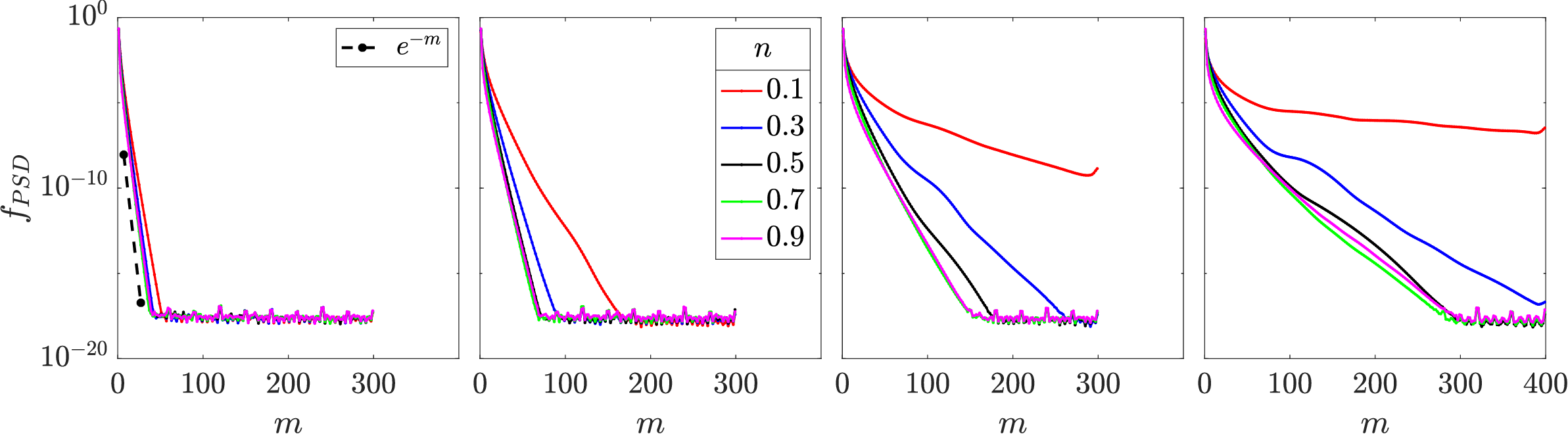} 
	\put(-420,105){$\Lambda=1$}\put(-303,105){$\Lambda=2$}\put(-187,105){$\Lambda=5$}\put(-60,105){$\Lambda=10$}
	\caption{Power spectral density $f_{PSD}=\sqrt{\int_{-h}^h |\hat{U}_{b,m}|^2 dy}$ of the laminar base flow $U_b$ in the Stokes layer of Carreau fluids at various power indices $n$ and Carreau numbers $\Lambda$. The dashed line in the first panel indicates a power law relation between the PSD and $m$. }
	\label{Fig:PSD_base_flow}
\end{figure}

\begin{figure}
	\centering
	\includegraphics[width=0.99\textwidth,trim= 0 0 0 0,clip]{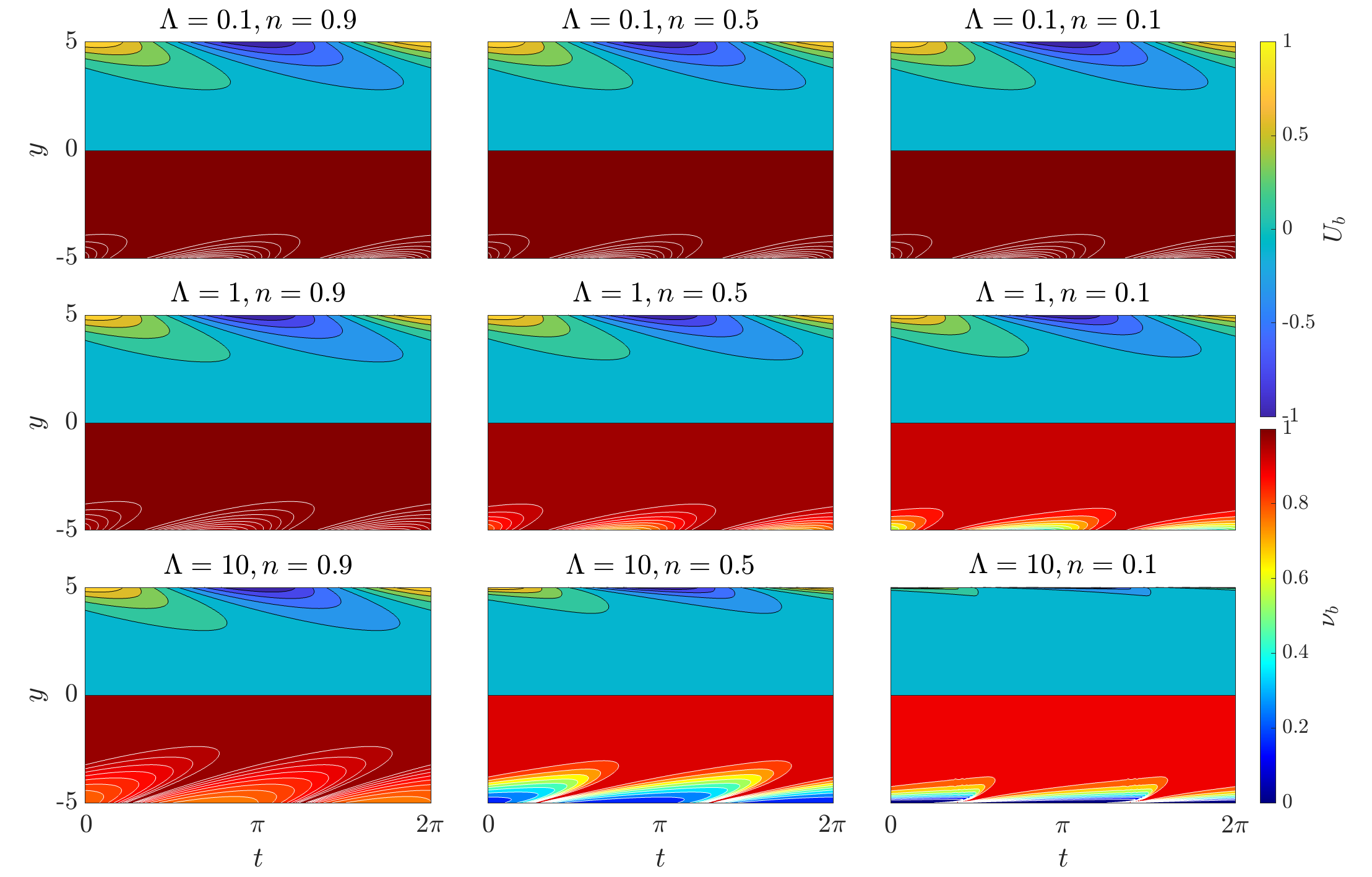}
	\caption{Variations of the velocity distribution $U_b(y,t)$ (top) and viscosity distribution $\nu_{b}(y,t)$ (bottom) in the laminar base flow of the Stokes layer with the power-law index $n$ and the Carreau number $\Lambda$. $\nu_b<1$ indicates shearing thinning.}
	\label{Fig:base_flow_varying_nlambda}
\end{figure}

To understand the relative strength of the harmonics in the base flow, the power spectral density (PSD) of the Fourier components at various $n$ and $\Lambda$ is shown in figure \ref{Fig:PSD_base_flow}. For relatively small $\Lambda=1$, the PSD of the lower-rank modes presents a scaling of $\sim e^{-m}$ approximately, with higher-rank modes being negligible. However, with smaller $n$ and larger $\Lambda$ (stronger shear-thinning effect), the PSD increases across a broader range of modes, meaning that more Fourier modes are needed to resolve accurately the time dependence of the flow. Notably, when $\Lambda=10$, at least about 300 modes are needed to accurately represent the base flow, which implies a substantially large matrix in the Floquet analysis and an exceedingly long time to conduct the analysis. This indicates the strong nonlinear effect at large $\Lambda$. Consequently, it becomes challenging to accurately determine the critical $Re_c$ in the large-$\Lambda$ limit, to be discussed shortly. 

It is of interest to probe how the parameters $n$ and $\Lambda$ change the base flow profiles. Figure \ref{Fig:base_flow_varying_nlambda} presents the velocity and viscosity distributions in the $y$-$t$ plane with varying $n=(0.9,0.5,0.1)$ and $\Lambda=(0.1,1,10)$. We find that by enhancing the shear-thinning effects the pattern in the  $y$-$t$ diagram of the base velocity profile gets steeper and the flow structures become increasingly localized near the walls. Correspondingly, the fluid viscosity within the Stokes layer thickness decreases much more. The effect of $n$ is more significant for larger $\Lambda$, see the results in rows for different $\Lambda$, with small $n$ leading to more localized structure. The localization and thus the sharp gradient of the base flow necessitate denser nodes at the wall and also more Fourier modes to resolve the temporal evolution, which is consistent with the result in figure \ref{Fig:PSD_base_flow}.

\subsection{Comparison between two methods for the base-flow solutions}

\begin{figure}
	\centering
	\includegraphics[width=0.99\textwidth,trim= 0 0 0 0,clip]{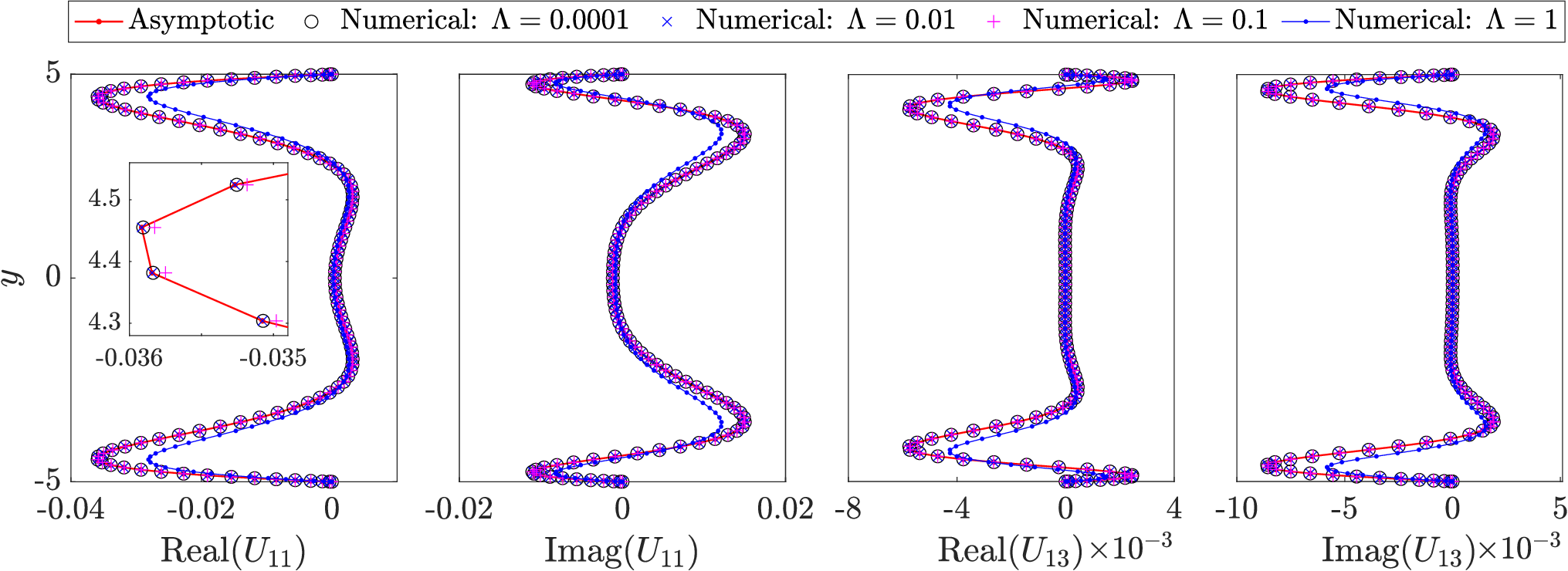} 
	\caption{Comparisons of the profiles of the leading harmonic $U_{11}$ and the third harmonic $U_{13}$ at various Carreau numbers $\Lambda$, obtained from the 2-terms expansion method ($\Ubz + \Lambda^2  \Ubo$) and the numerical method ($U_{b,num}$). The power-law index is fixed at $n=0.5$. The inset in the leftmost panel shows an enlarged view of the panel. Asymptotic here means the expanded base flow to be compared with an asymptotically small $\Lambda$ in the numerical base flow. }
	\label{Fig:base_flow_compare}
\end{figure}

\begin{figure}
	\centering
	\includegraphics[width=0.75\textwidth,trim= 0 0 0 0,clip]{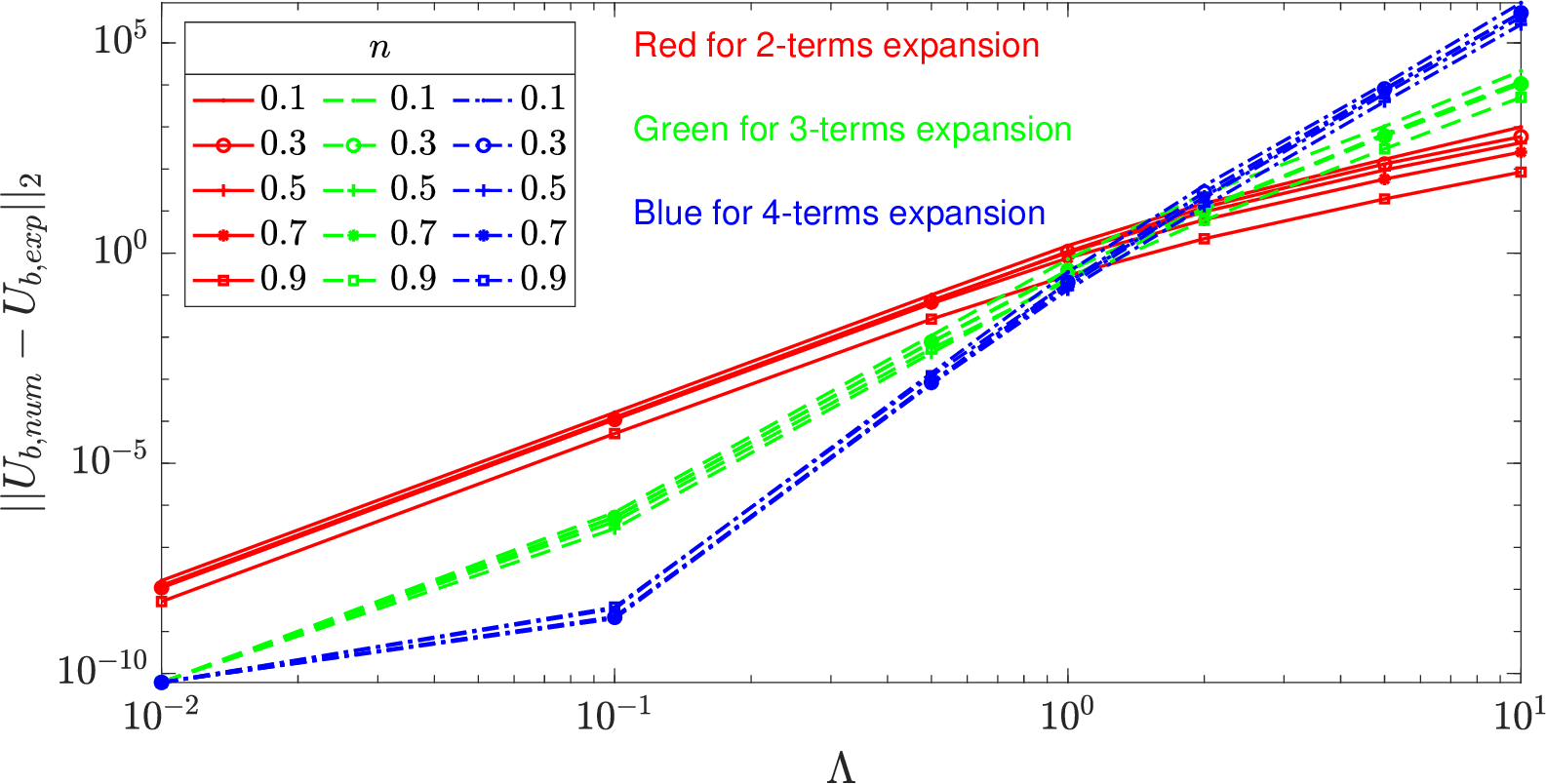}
	\caption{$L_2$-norm of the difference between the base flow velocities obtained using the  the numerical method $U_{b,num}$ and the expansion method $U_{b,exp}$ based on 2-term ($\Ubz + \Lambda^2  \Ubo$), 3-term ($\Ubz + \Lambda^2  \Ubo  + \Lambda^4   \Ubt$), 4-term ($\Ubz + \Lambda^2  \Ubo  + \Lambda^4   \Ubt+ \Lambda^6   \Ubth$) expansions as in Eq. \eqref{baseflowexpand}.}
	\label{Fig:base_flow_compare_moreterms}
\end{figure}

The base flow profiles can be Fourier-expanded as in Eq. \eqref{twoharmonicbase}. The first two harmonics of the obtained solutions using the numerical and expansion methods at $n=0.5$ are compared in figure \ref{Fig:base_flow_compare}; see equation \eqref{U11_U13} for the definition of the complex variables $U_{11}$ and $U_{13}$ in the expansion method. 
For small $\Lambda$ from 0.0001 to 0.1, the expanded solution matches well with the reference numerical solution. Detailed examination of the results reveals a slight deviation as shown in the inset for $\Lambda=0.1$, which is acceptable to the graphical accuracy. But large discrepancy can be clearly seen at $\Lambda=1$. This comparison aims to demonstrate the validity of the expansion method at small $\Lambda$ and also verifies the fully spectral method. 

To provide a comprehensive assessment of the discrepancy between the two methods, the $L_2$-norm of their difference is computed for various truncation orders in the expansion method, as shown in Figure~\ref{Fig:base_flow_compare_moreterms}. The results demonstrate that incorporating additional terms in the expansion significantly improves accuracy for $\Lambda \le 1$, across all power-law indices $n$, a trend consistent with the expected behavior of asymptotic series expansions. \dongdong{It seems to be an exception that at $\Lambda=0.01$, the results based on 4-terms expansion yield the same error as those based on 3-terms expansion. As suggested by one of the reviewer, it is probably due to a lack of accuracy in the numerical base flow solution $U_{b,num}$. To check this, we have used much higher resolutions to re-calculate $U_{b,num}$, but no further improvement can be obtained. Therefore, we suspect the machine error prevents us from getting more accurate numerical solutions. } For larger values of $\Lambda$, the accuracy of the expansion method deteriorates compared to the numerical solution, suggesting that the series has a convergence radius of approximately one.

Having shown that the expansion method can accurately reproduce the numerical base flow for small $\Lambda$, we further demonstrate in Table~\ref{Tab:comparison_eigenvalue} that the leading eigenvalues from the Floquet analysis, based on these two base flows, are also in close agreement, provided the value of $\Lambda$ is small. Specifically, for $\Lambda\le 0.2$, the relative error of the expansion method is consistently less than 1\% compared to the numerical method, and incorporating higher order terms in the expansion method increases the accuracy. This indicates that the expansion base flow can accurately capture the linear dynamics of the Stokes layer flow of Carreau fluids with weak shear-thinning effects. Since this solution is analytical, it offers potential advantages for theoretical analysis in the small $\Lambda$ limit. On the other hand, for $\Lambda  \geq \mathcal O(1)$, the expansion method introduces increasing inaccuracies; incorporate higher order terms in fact deteriorates the results. Therefore, in the rest of the paper, for cases with $\Lambda  \geq \mathcal O(1)$ (intermediate and strong shear-thinning), we will rely on the numerical base flow for all the eigenvalue problem calculations.

\begin{table}
	\begin{ruledtabular} 
			\begin{tabular}{p{0.7cm}p{2.5cm}p{3.0cm}p{3cm}p{3cm}}
		$\Lambda$     &  Using numerical base flow  & Using 2-terms expansion base flow  & Using 3-terms expansion base flow & Using 4-terms expansion base flow \\ \hline
		$0.0001$ &   $0.00106691$  & $0.00106691 (0.000\%)$ & $0.00106691 (0.000\%)$ & $0.00106691 (0.000\%)$ \\
		$0.01$   &     $0.00101669$  & $0.00101669 (0.000\%)$ & $0.00101669 (0.000\%)$ & $0.00101669 (0.000\%)$ \\
		$0.1$ &         $-0.00396355$ & $- 0.00397137 (0.197\%)$ & $-0.00396381 (0.007\%)$ & $-0.00396355 (0.000\%)$ \\
		$0.2$ &         $-0.0191380$ & $-0.0192478 (0.574\%)$ & $-0.0191546 (0.087\%)$ & $-0.0191374 (-0.003\%)$ \\
		$0.5$ &   $-0.126170$ & $-0.127315 (0.908\%)$ & $-0.130021  (3.05\%)$ & $-0.125462 (-0.561\%)$ \\
		$1$ &  $-0.445534$ & $-0.404360 (-9\%)$ &$-0.619105 (39\%)$ & $-0.326863 (-26\%)$ \\
		$2$ &  $-0.907486$& $-0.422183 (-53\%)$ &$-0.545266 (-40\%)$ & $-3.40225 (275\%)$ \\
			\end{tabular} 
		\caption{Comparison of the linear growth rate of the most unstable or least stable mode obtained using two different calculation methods for the base state at varying $\Lambda$ and $(n,Re,\alpha)=(0.5, 648, 0.41)$. The resolution is set at $N_y=99$ and $N_f=300$. The percentage in the parenthesis indicates the relative error with respect to the reference numerical method.} 
		\label{Tab:comparison_eigenvalue}
	\end{ruledtabular}
\end{table}

\subsection{Effect of various parameters $\Lambda, n, \Omega$}

\begin{figure}
	\centering
	\includegraphics[width=0.33\textwidth,trim= 0 0 0 0,clip]{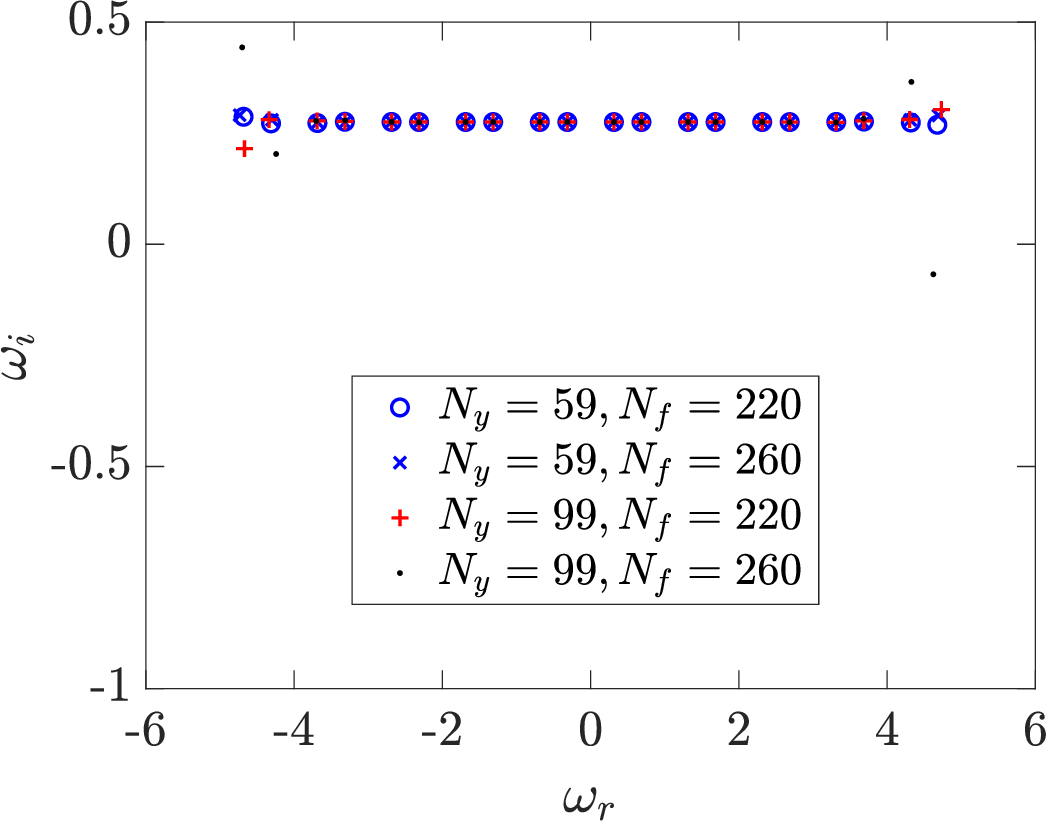}\put(-170,132){$(a)$}
	\includegraphics[width=0.33\textwidth,trim= 0 0 0 0,clip]{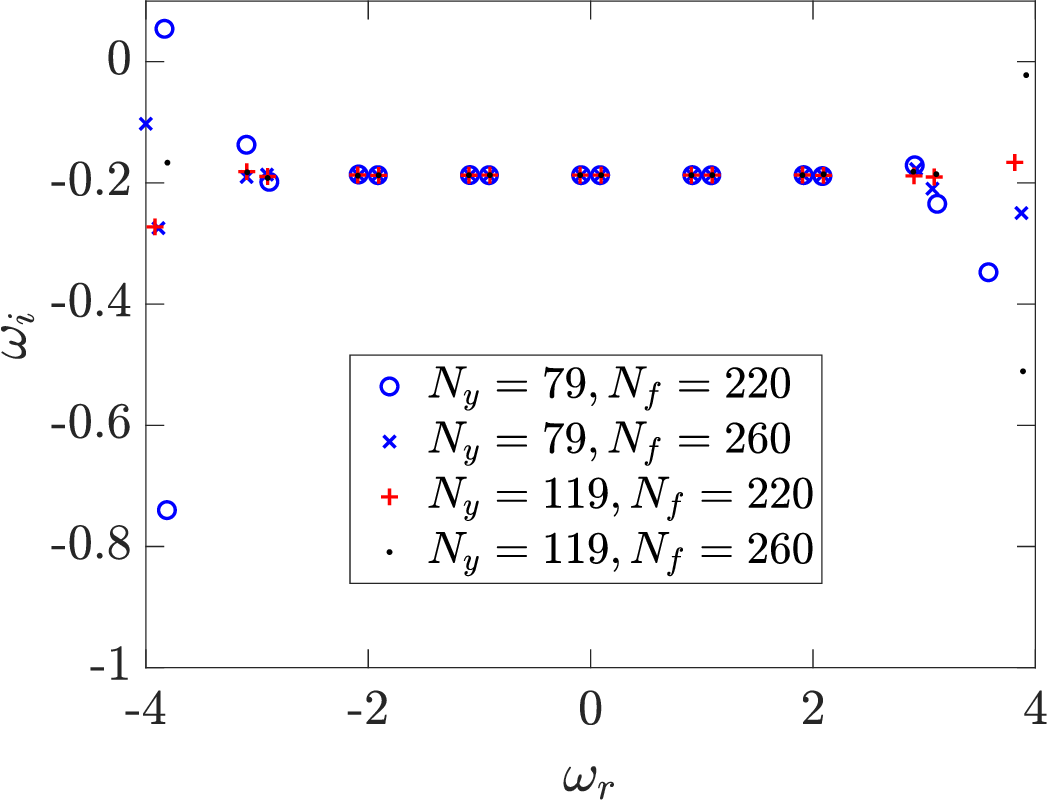}\put(-167,132){$(b)$}
	\includegraphics[width=0.33\textwidth,trim= 0 0 0 0,clip]{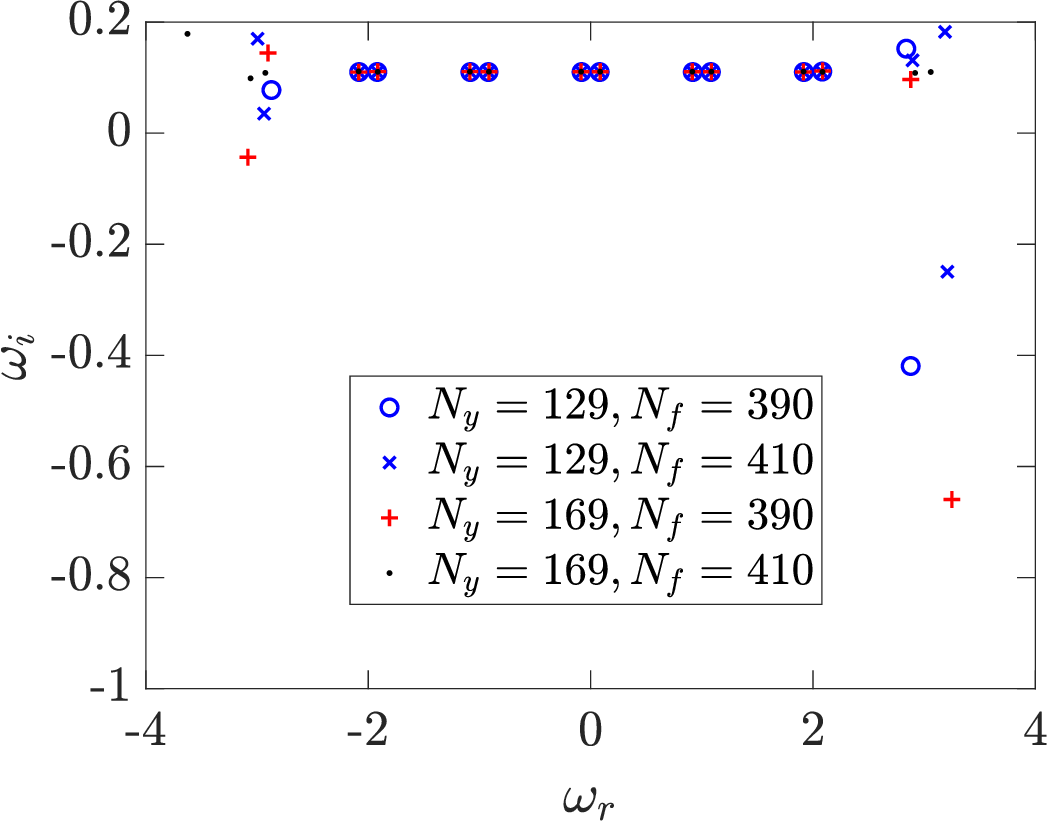}\put(-169,132){$(c)$}
	\caption{ \color{black} Eigenspectra of the Stokes layer flow  of Carreau fluids in a channel at $Re=700$ and $(a)$ $(n,\Lambda,\alpha)=(0.9, 0.1, 0.41)$, $(b)$ $(n,\Lambda,\alpha)=(0.5, 1, 0.41)$, $(c)$ $(n,\Lambda,\alpha)=(0.5, 8, 0.7)$. $\omega_i$ and $\omega_r$ are the linear growth rate and frequency of the eigenmode, respectively. $\omega_{i}>0$ indicates linear instability. }
	\label{Fig:eigenspectra}
\end{figure}

{\color{black}
FIG. \ref{Fig:eigenspectra} shows the eigenspectra of the flow at different parameter settings but with a fixed Reynolds number of $Re=700$. In each of them, four sets of resolution settings are used to demonstrate the convergence of the eigenspectra; some data are listed in Appendix \ref{convergence_appendix}. As one can see, the convergence can be attained best for the Floquet modes at around zero frequency $\omega_r=0$ and it deteriorates when $|\omega_r|$ becomes larger. This behavior is reasonable. As evident from Eq. \eqref{eq:v}, for any eigenvalue $\omega$ and its complex conjugate $\omega^*$, $\omega+m$ and $\omega^*+m$ are also eigenvalues of the same system for any integer $m$, suggesting that theoretically there should be infinite number of eigenvalues with different frequencies but exactly the same linear growth rate. Numerically, it is impossible and unnecessary to directly obtain all of them in the eigenspectra. In practice, we solve for those few most unstable/least stable eigenmodes with the lowest $|\omega_r|$. For example, to obtain FIG. \ref{Fig:eigenspectra}$(a)$ the eigs() function in MATLAB is employed and only the leading 20 eigenmodes closest to the initial guess $0+1i$ are solved. Therefore, the modes far from the initial guess have worse convergence. Nevertheless, we always and just need to focus on the eigenmodes with frequencies in $[0, 1/2]$ under the current linear analysis framework; see, e.g., Ref. \cite{Blennerhassett2006}. Regarding the accuracy of the eigenmodes, the four modes with the lowest $|\omega_r|$ in FIG. \ref{Fig:eigenspectra}$(a)$ have eigenvalues: $0.314363+0.275047i$, $0.685636+0.275046i$ and their complex conjugates.  Correspondingly, we have checked that the eigenfunctions of a pair of complex conjugate modes have half of a period, i.e., $\pi$ shift in time $t$, suggesting perturbations propagating both in the positive and negative $x$ directions with the same growth rate. Such consistency also serves as a validation of our computation.
}

\begin{figure}
	\centering
	\includegraphics[width=0.48\textwidth,trim= 0 0 0 0,clip]{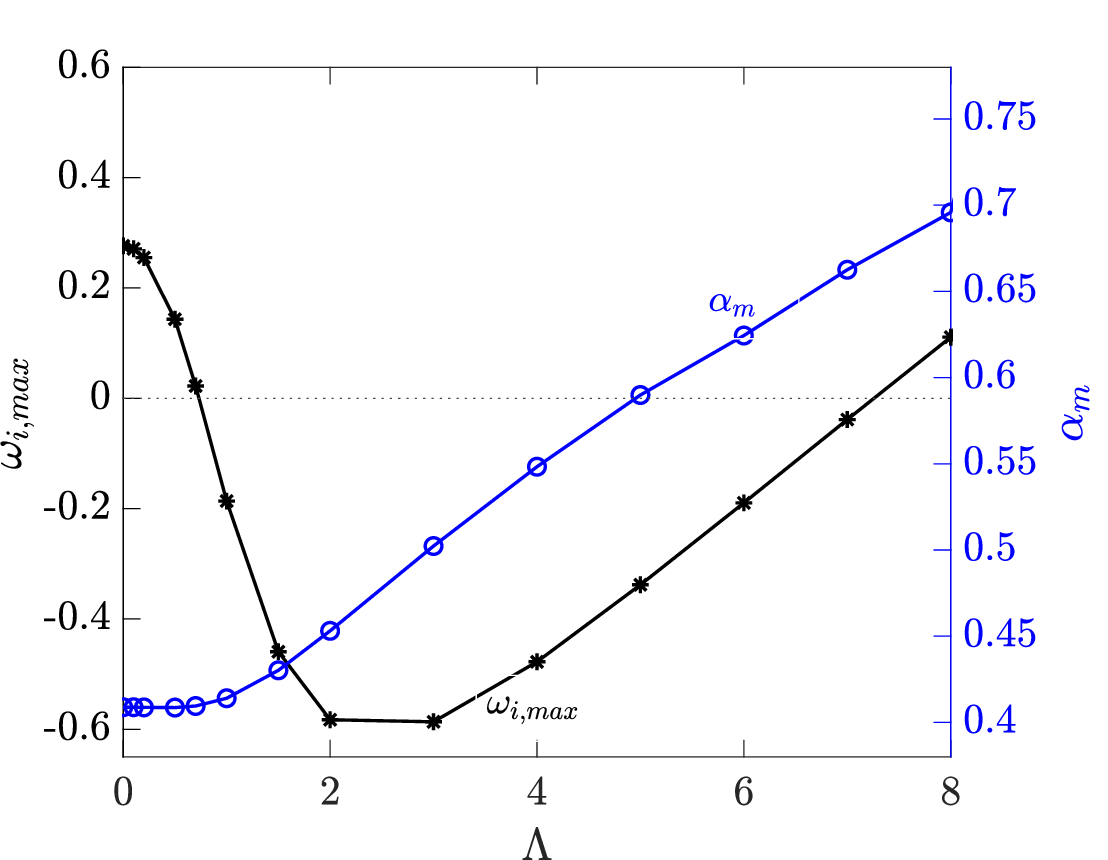}\put(-247,180){$(a)$}
	\includegraphics[width=0.48\textwidth,trim= 0 0 0 0,clip]{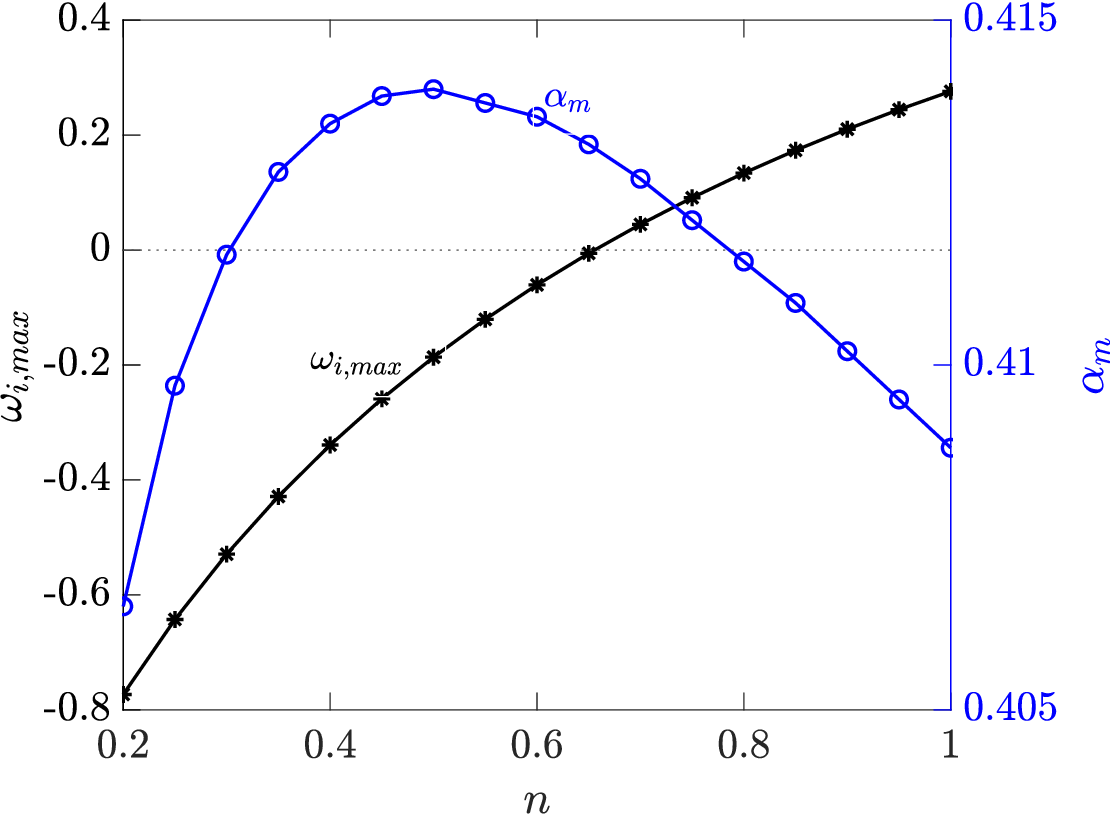}\put(-247,180){$(b)$}
	\caption{ \color{black} The maximal linear growth rate $\omega_{i,max}$  corresponding to the most unstable or least stable mode attained at the wavenumber $\alpha_m$ in the Stokes layer for $(a)$ varying Carreau number $\Lambda$ with $(n,Re)=(0.5, 700)$ and $(b)$ varying power-law index $n$ with $(\Lambda,Re)=(1, 700)$. $\omega_{i,max}>0$ indicates linear instability. }
	\label{Fig:GR_FR_varying_lambda}
\end{figure}

\dongdong{To investigate the effect of shear-thinning on the oscillatory flow stability, we show in Figure \ref{Fig:GR_FR_varying_lambda}$(a)$ the variation of the maximal linear growth rate $\omega_{i,max}$ (maximized over the wavenumber and attained at $\alpha_m$) for $n = 0.5$ and $Re = 700$. The results reveal a non-monotonic effect of increasing $\Lambda$. Specifically, strengthening the shear-thinning effect from zero (Newtonian) to an intermediate level (at $\Lambda \approx 3$) stabilizes the flow. For weakly shear-thinning fluids, Sun et al. \cite{Sun2017Flow} found experimentally that shear-thinning enhanced the flow stability in a microfluidic oscillator. Further increasing the shear-thinning effect in our flow (from $\Lambda \approx 3$ to $\Lambda = 8$), however, destabilizes the flow. The flow becomes linearly unstable when $\Lambda \gtrsim 7.2$. }
Such a non-monotonic effect of shear-thinning has also been reported by Nouar \textit{et al}.~\cite{Nouar2007Delaying} in their study of plane Poiseuille flow of Carreau fluids, suggesting that shear-thinning may exert generic influences on parallel shear flows despite differences in the temporal characteristics of their laminar base states. A more detailed discussion on the stabilising/destabilising effect of $\Lambda,n$ will be provided in the next section, where we focus on the critical value of $Re$.

In contrast to the non-monotonic effect of $\Lambda$, reducing $n$ is found to be monotonically stabilizing as can be seen from Figure \ref{Fig:GR_FR_varying_lambda}$(b)$. This effect of $n$ is also consistent with that observed in \cite{Nouar2007Delaying} for plane Poiseuille flows of Carreau fluids. 

To facilitate comparisons with future experimental studies on this shear-thinning instability, we consider the variation of the dimensional oscillation frequency $\Omega$, which is the most practical way to vary in experiments. In Newtonian Stokes layer flows, experimental studies \cite{Merkli1975Transition} and \cite{Hino1976Experiments} have reported the effect of $\Omega$ in purely oscillatory pipe Stokes layers and oscillating flows in tubes, respectively. In the shear-thinning flow, we found that solely varying $\Omega$ exhibits a non-monotonic effect on the flow stability/instability as presented below.

To illustrate this effect, we select the parametric point at $h_{\text{ref}}=5$, $n=0.9$, $\Lambda_{\text{ref}}=1$ and $Re_{\text{ref}}=660$ as the reference data point; its corresponding dimensional frequency is denoted as $\Omega_{\text{ref}}$. Then we define the frequency ratio as
\begin{equation}
r_{\omega} = \frac{\Omega}{\Omega_{\text{ref}}}.
\end{equation}
With all the other dimensional control parameters being fixed and changing $\Omega$ only, the dimensionless parameters defined in the present study vary with $r_{\omega} $ as
\begin{equation}\label{para_vary_omega}
h = h_{\text{ref}}\sqrt{r_{\omega} }, \, \, \, \Lambda=\Lambda_{\text{ref}}\sqrt{r_{\omega} },  \, \, \, Re = \frac{Re_{\text{ref}}}{\sqrt{r_{\omega} }}.
\end{equation}

\begin{figure}
	\centering
	\includegraphics[width=0.48\textwidth,trim= 0 0 0 0,clip]{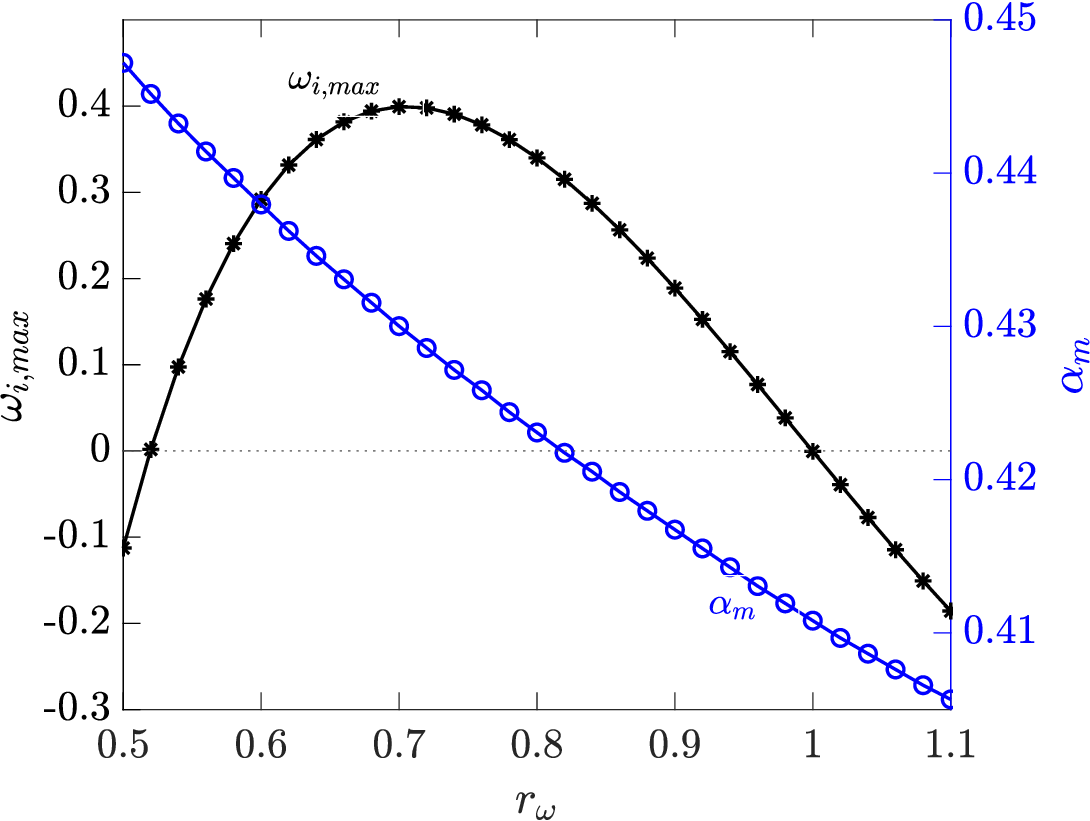}
	\caption{\color{black} The maximal linear growth rate $\omega_{i,max}$ corresponding to the most unstable or least stable mode attained at the wavenumber $\alpha_m$ in the Stokes layer at varying oscillation frequency ratios $r_{\omega}$. The reference parametric point is  $h_{\text{ref}}=5$, $n=0.9$, $\Lambda_{\text{ref}}=1$ and $Re_{\text{ref}}=660$; note that $(h,\Lambda,Re)$ change with $r_{\omega}$ according to equations \eqref{para_vary_omega};  $\omega_{i,max}>0$ indicates linear instability. }
	\label{Fig:GR_FR_varying_omega}
\end{figure}

Figure \ref{Fig:GR_FR_varying_omega} shows that starting from \dongdong{ the above reference parametric point $r_w=1$ } which is approximately neutral, altering the oscillation frequency has a non-monotonic effect on the flow stability/instability. The linear growth rate is positive within a specific frequency range and the trend suggests that the flow is linearly stable at both vary low and very high frequencies. The corresponding leading modes are all oscillatory (not shown), traveling in the streamwise direction in each cycle.

%The wavenumber is changing in figure \ref{Fig:GR_FR_varying_omega}($a$), according to equation \eqref{para_vary_omega}. We realise that when an experiment or a numerical simulation is conducted, the streamwise length is fixed, implying that a dominant wavenumber is fixed. In view of this, we also fixed $\alpha$ (to be equal to 0.41) to explore the effect of oscillation frequency $\omega$. The non-monotonic effect of varying the oscillation frequency persists in this case, as can be seen from  figure \ref{Fig:GR_FR_varying_omega}$(b)$.

As a related result, Zhang \cite{Zhang2025} conducted the transient growth analysis of the Newtonian Stokes layer with varying $\Omega$ and also found that the maximum energy perturbation is a non-monotonic function of $\Omega$, which is consistent with our results of the Floquet analysis. In the literature of other periodic complex fluids, Ref. \cite{Casanellas2014Vortex} reported analogous non-monotonic effect of oscillation frequency in their experimental study on purely oscillatory pipe flow of wormlike micelle solutions fluids, which was later corroborated theoretically by \cite{Ortin2020}. Despite the differences in the flow configurations, the common effect of oscillation frequency suggests possible generic characteristics of Stokes layer flows of complex fluids, that is, too small or too large oscillating frequencies stabilise the flow, whereas, a certain range of the frequencies destabilises the flow, which is characteristically different in different flows.

These results suggest a way for controlling flow instability in the finite Stokes layer flow. On one hand, the most violent instability can be triggered at an optimal oscillation frequency of an intermediate level, which may be employed for enhancing flow mixing. On the other hand, one may consider using either a very low or a very high oscillation frequency to suppress the possible linear instability in the flow. Nevertheless, it should be noted that this does not necessarily guarantee a smooth laminar flow, because nonlinear instabilities cannot be ruled out which calls for further explorations in an experimentally feasible manner.

\subsection{Neutral curves}

\begin{figure}
	\centering
	\includegraphics[width=0.99\textwidth,trim= 0 0 0 0,clip]{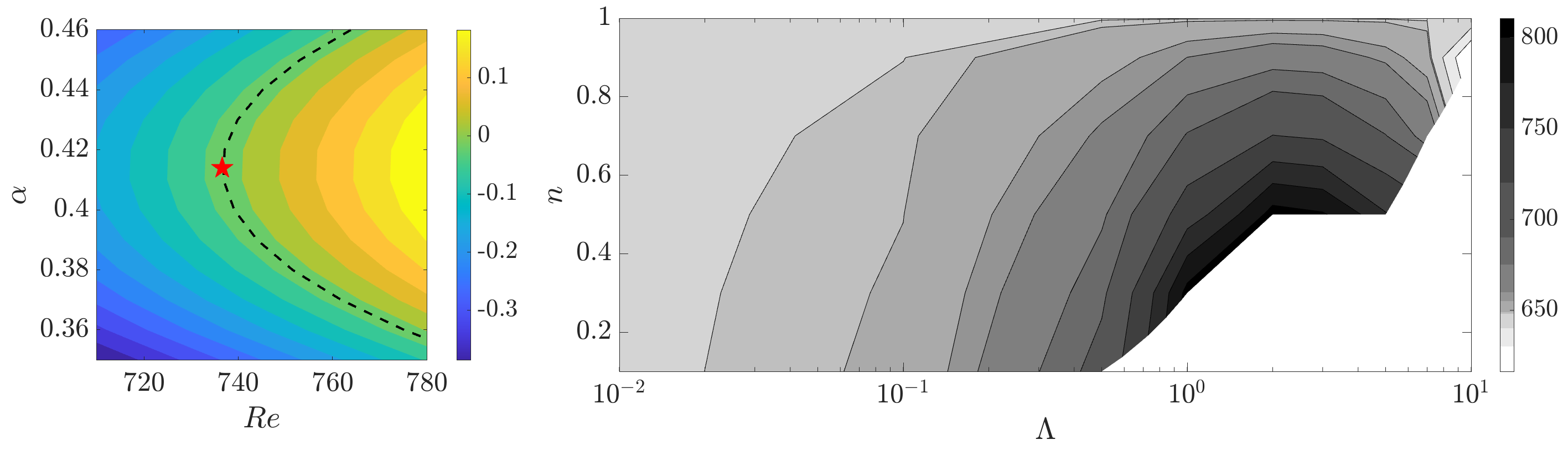}
	\put(-505,135){$(a)$}\put(-332,135){$(b)$}\\
	\includegraphics[width=0.59\textwidth,trim= 0 0 0 0,clip]{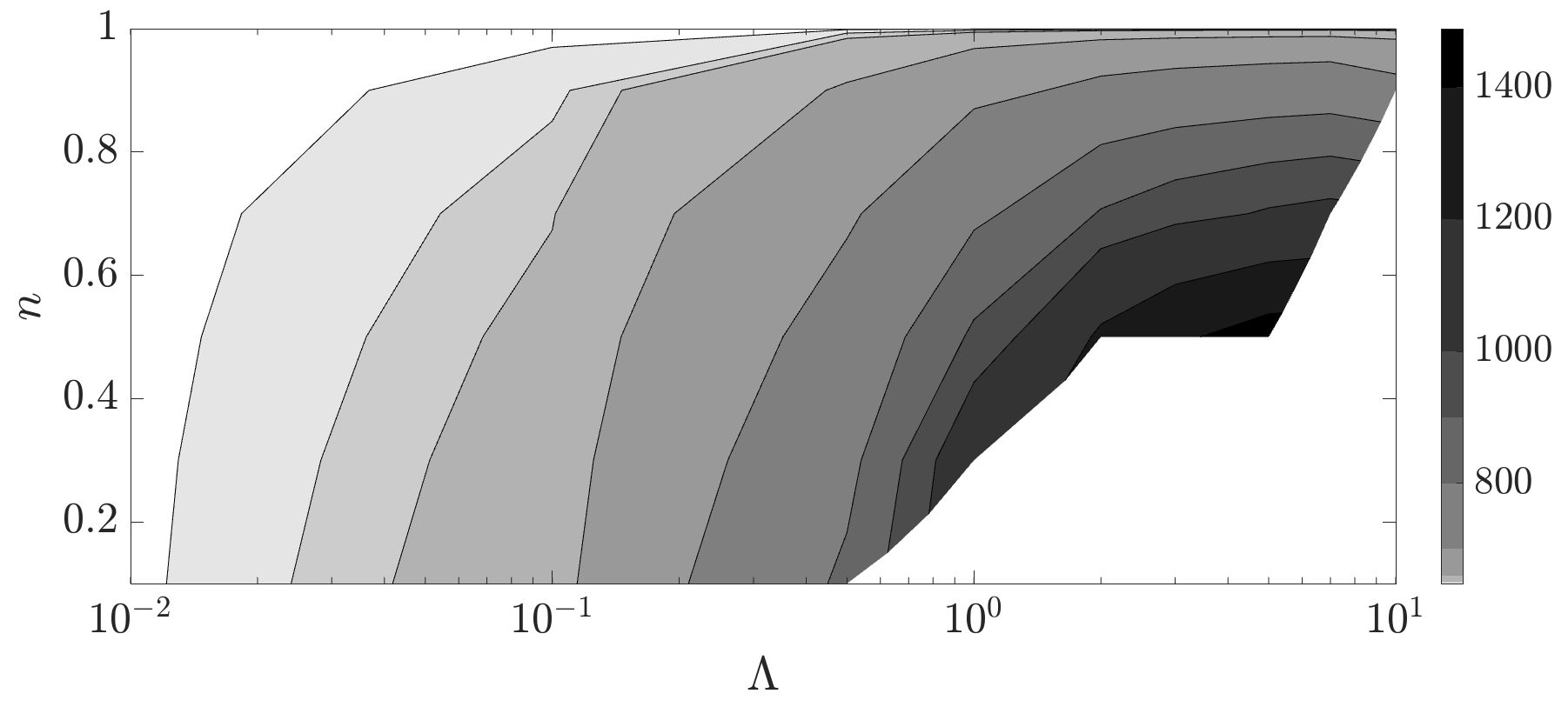}
	\put(-300,127){$(c)$}
	\caption{$(a)$ Contours of the linear growth rate $\omega_i$ and the neutral curve (black dashed line along which $\omega_i=0$) for the Stokes layer of Carreau fluids at $n=0.5$ and $\Lambda=1$. The red star marks the linear critical condition $Re_c\approx736.66$, $\alpha_c\approx0.414$.  $(b)$ Contour of the linear critical Reynolds number $Re_c$ in the $\Lambda$-$n$ plane. $(c)$ Contour of the linear critical Reynolds number $Re_{t,c}$ based on the time-averaged tangent viscosity at the walls. }
	\label{Fig:NC_Rec_alpc_varying_nlambda}
\end{figure}

By varying $Re$ and $\alpha$ at given $n$ and $\Lambda$, we can plot the contours of the linear growth rate $\omega_i$ in the $Re$-$\alpha$ plane; see figure \ref{Fig:NC_Rec_alpc_varying_nlambda}$(a)$ for such an example at $n=0.5$ and $\Lambda=1$. The dashed curve in the figure indicates $\omega_i=0$ and is known as the neutral curve. Along the neutral curve, the minimal $Re$, marked by the red star, is called the linear critical Reynolds number $Re_c$, corresponding to the linear critical wavenumber $\alpha_c$. For the present case at $n=0.5$ and $\Lambda=1$, $Re_c\approx 736.66$ and $\alpha_c \approx 0.414$; this critical Reynolds number is higher than that of the Newtonian counterpart at $Re_{c,Newtonian}\approx 647.80$ for $h=5$, while the critical wavenumber is only slightly larger than $\alpha_{c,Newtonian}\approx 0.410$. The increased $Re_c$ indicates the stabilizing effect of the weakly shear-thinning Carreau fluids.

The critical Reynolds number $Re_c$ in the parameter space of $\Lambda$ and $n$ is computed and presented in Figure~\ref{Fig:NC_Rec_alpc_varying_nlambda}$(b)$. It should be acknowledged that accurately determining $Re_c$ becomes exceedingly challenging for large $\Lambda$ and small $n$, corresponding to a strong shear-thinning regime. As explained in the previous section, this difficulty arises from the need for a large number of modes to resolve both the highly nonlinear shear-thinning base flow and the associated eigenvalue problem. The regions where convergence cannot be reliably achieved are left blank in the figure. Despite these limitations, the available data still reveal meaningful qualitative trends. For large $n$ (i.e., weak shear-thinning), increasing $\Lambda$ initially raises $Re_c$, followed by a decline. For instance, at $n = 0.9$ and $\Lambda = 10$, $Re_c$ drops to 616.26, which is even lower than the Newtonian value of 647.80, demonstrating the destabilizing effect of strong shear-thinning Carreau fluids on Stokes layer flow. Conversely, for small $n$, the current data indicate a monotonic increase in $Re_c$ up to $\Lambda \sim \mathcal{O}(1)$. As seen in the figure, it is reasonable to extrapolate that $Re_c$ may eventually decrease again for larger $\Lambda$. The global maximum of $Re_c$ appears to occur around $\Lambda \sim \mathcal{O}(1)$.

The non-monotonic variation of \( Re_c \) with \( \Lambda \) has also been reported in Nouar et al. \cite{Nouar2007Delaying} for plane Poiseuille flow of Carreau fluids (where the shear-thinning effect is quantified as \( \lambda \)). Notably, Ref. \cite{Nouar2007Delaying} extended their analysis with a physical argument, that is, the disturbance field will most likely be influenced by the tangent viscosity, rather than the viscosity averaged across the channel. This led them to define a Reynolds number using the tangent viscosity at the channel walls. They found that in this case shear-thinning has a consistently stabilizing effect (compared to the Newtonian counterpart), delaying the transition to turbulence. To test this definition of $Re$ in the time-periodic flow, we intend to define a similar $Re$; however, the wall-based definition cannot be directly applied in the present study, as the tangent viscosity in Stokes layer flow is time-dependent. Instead, we introduce a modified Reynolds number \( Re_{t,c}=\frac{U_0}{\sqrt{2\overline{\nu_{b_t}}\Omega}}  \) based on $\overline{\nu_{b_t}}$ which is the time-averaged tangent viscosity at the channel walls. Figure~\ref{Fig:NC_Rec_alpc_varying_nlambda}$(c)$ illustrates that the value of $Re_{t,c}$ increases monotonically within the investigated $\Lambda,n$ values (except a slight decrease in the region where $n$ and $\Lambda$ are both large), indicating that the shear-thinning effect becomes consistently stabilizing in general when the flow is characterized by the tangent-viscosity-based Reynolds number. More specifically, we also found that compared to the Newtonian Stokes layer, the shear-thinning has a stabilising effect. This finding is broadly consistent with \cite{Nouar2007Delaying}, despite the differences in flow configurations. We summarise the comparison of our results and those in Ref. \cite{Nouar2007Delaying} in table \ref{Re_Lambda_comparison}. 

\begin{table}[h!]
\centering
\caption{Comparison of Reynolds number and shear-thinning parameter definitions across different studies, and the resulting trends in stability. In Ref. \cite{Nouar2007Delaying}, $U_c$ is the laminar velocity at the channel center, $h$ denotes half-channel thickness, their dimensional $\hat \lambda$ is our $\lambda$ (which is also dimensional), $\overline{\hat \mu_0}$ represents the zero-shear-rate viscosity coefficient averaged over the channel and $\mu_{tw}$ stands for the wall tangent viscosity coefficient.}
\label{Re_Lambda_comparison}
\begin{tabular}{|c|c|c|p{9cm}|}
\hline
Works & $Re$ definition & Shear-thinning parameter & Conclusions \\
\hline
Nouar et al. \cite{Nouar2007Delaying} & 
$\overline{Re} = \dfrac{\rho U_{c}h}{\overline{\hat\mu_{0}}}$ & 
$\lambda = \dfrac{\hat\lambda}{h/U_{c}}$ & 
$\overline{Re}$ vs.\ $\lambda$: \textbf{non-monotonic}, see their figures~4a. Relative to Newtonian flow, large $\lambda$ is destabilising. \\
\hline
Nouar et al. \cite{Nouar2007Delaying} & 
$Re_{tw} = \dfrac{\rho U_{c}h}{\mu_{tw}\hat\mu_{0}}$ & 
$\lambda = \dfrac{\hat\lambda}{h/U_{c}}$ & 
$Re_{tw}$ vs.\ $\lambda$: \textbf{non-monotonic}, see their figure~5a. Relative to Newtonian flow, large $\lambda$ is stabilising. \\
\hline
Present work & 
$Re = \dfrac{U_{0}}{\sqrt{2\nu_{0}\Omega}}$ & 
$\Lambda = \dfrac{\lambda}{\sqrt{2\nu_{0}/\Omega}\, U_{0}}$ & 
$Re$ vs.\ $\Lambda$: \textbf{non-monotonic}, see our figure~8b. Relative to Newtonian flow, large $\lambda$ is destabilising.  \\
\hline
Present work & 
$Re_{t} = \dfrac{U_{0}}{\sqrt{2\overline{\nu_{bt}}\Omega}}$ & 
$\Lambda = \dfrac{\lambda}{\sqrt{2\nu_{0}/\Omega}\, U_{0}}$ & 
$Re_{t}$ vs.\ $\Lambda$: mostly monotonic except for large $n,\Lambda$; see our figure~8c. If larger $\Lambda$ were explored, the trend would likely become \textbf{non-monotonic} as well \\
\hline
\end{tabular}
\end{table}

%Specifically, in shear-thinning Poiseuille flow, the dependence of the Reynolds number on~$\lambda$ is likewise non-monotonic: see figures~3a and~4a of~\cite{Nouar2007Delaying} when the Reynolds number is defined using the viscosity averaged across the channel, and figure~5a when it is defined in terms of the wall-tangent viscosity. Despite these similar observations, we emphasise that the two flow configurations are fundamentally different, and our result represents a new finding rather than a recovery of a previously published conclusion. In later sections, we will further elucidate the instability mechanism in the time-dependent flow, which stands in contrast to that of the time-independent configuration.

\subsection{Destabilizing mechanism}\label{sec:destabilizing_mechanism}
In the previous two sections, we revealed the destabilizing effects at both small and large values of $\Lambda$. This section aims to further elucidate the nature of the identified oscillatory flow instability through an energy analysis. In the context of shear-thinning instabilities, Nouar et al. \cite{NOUAR2007} performed an energy analysis for a steady channel flow. In contrast, our energy analysis is conducted on a time-periodic base flow. We find that the phase difference between the disturbance field and the oscillatory laminar base flow plays a central role in determining the stability or instability of the system.

To conduct the energy analysis, the linearized perturbation equation \eqref{linearizedeq} is multiplied by the perturbation itself. To facilitate the discussion, we will denote the shape function $\Big[\sum_{m=-\infty}^\infty {v}^{(m)}(y) e^{imt}\Big]$ in Eq. \eqref{solutionform} as $\tilde {v}(y,t)$ in this section.
The perturbation kinetic energy density for the shape function is then defined as $E_k(y,t)=\frac{1}{2} (\tilde{u}\tilde{u}^* + \tilde{v}\tilde{v}^* )$, whose time derivative is
\begin{equation}
\frac{\partial E_k}{\partial t} = \frac{1}{2} \left(\frac{\partial \tilde{u}}{\partial t} \tilde{u}^* +\frac{\partial \tilde{v}}{\partial t} \tilde{v}^*  \right)+ \text{c.c.}
\end{equation}
Using Eq. (\ref{linearizedeq}) leads to
\begin{align}\label{energy_density}
2\omega_i E_k + \frac{\partial E_k}{\partial t} = & \underbrace{\left[-\frac{Re}{2} \tilde{u}^* \tilde{v}\frac{\partial U_b}{\partial y} + \text{c.c.}\right]}_{\text{Production $f_{PR}$}} +   \underbrace{\left[-\frac{Re}{2} \left( i\alpha \tilde{u}^*\tilde{p} + \tilde{v}^* \frac{\partial \tilde{p}}{\partial y} \right) + \text{c.c.}\right]}_{\text{Pressure  work $f_{PW}$}} \notag\\ 
%%%%
& + \underbrace{\frac{1}{4} \nu_b (\tilde{u}^* \tilde{\nabla}^2 \tilde{u} + \tilde{v}^* \tilde{\nabla}^2 \tilde{v}   ) + \text{c.c.}}_{\text{Viscous dissipation $f_{VD}$}} +  \underbrace{\frac{1}{4} \frac{\partial \nu_{bt}}{\partial y} (\tilde{u}^* \frac{\partial \tilde{u}}{\partial y} + i\alpha \tilde{u}^* \tilde{v} ) +  \frac{1}{4} \frac{\partial \nu_{b}}{\partial y} (2 \tilde{v}^* \frac{\partial \tilde{v}}{\partial y}  ) + \text{c.c.}}_{\text{Additional work $f_{VS}$}} \notag\\ 
& +  \underbrace{\frac{1}{4} (\nu_{bt}-\nu_b) \left(\tilde{u}^* \frac{\partial^2 \tilde{u}}{\partial y^2} + i\alpha  \tilde{u}^* \frac{\partial \tilde{v}}{\partial y}  + i\alpha \tilde{v}^* \frac{\partial \tilde{u}}{\partial y}  - 2\alpha^2\tilde{v}^* \tilde{v} \right) + \text{c.c.}}_{\text{Additional dissipation $f_{VT}$}} 
\end{align} 
On the right hand side, the underscored five terms from top to bottom are respectively: the production term $f_{PR}$, the pressure work $f_{PW}$, the viscous dissipation $f_{VD}$, the additional work due to viscosity stratification  $f_{VS}$, and finally the additional dissipation due to the viscosity difference between $\nu_{bt}$ and $\nu_b$, denoted as $f_{VT}$. 

We will investigate the spatial-temporal distribution of these quantities and examine how they contribute to the distribution of the perturbation energy density $E_k$. To analyze the system as a whole, we integrate Eq. \eqref{energy_density} in the wall-normal direction and in time, resulting in
\begin{align} \label{eq4d3}
2\omega_i \bar{\bar{E}}_{k} = \bar{\bar{f}}_{PR} +\bar{\bar{f}}_{VD} +\bar{\bar{f}}_{VS}  + \bar{\bar{f}}_{VT} \ \ \ \ \text{or} \ \ \ \ 2\omega_i  = \frac{\bar{\bar{f}}_{PR}}{\bar{\bar{E}}_{k}} + \frac{\bar{\bar{f}}_{VD}}{\bar{\bar{E}}_{k}} + \frac{\bar{\bar{f}}_{VS}}{\bar{\bar{E}}_{k}}  + \frac{\bar{\bar{f}}_{VT} }{\bar{\bar{E}}_{k}}.
\end{align}
Here, the double overbar represents the operation $\bar{\bar{f}} = \frac{1}{2\pi} \frac{1}{2h}\int_0^{2\pi} \int_{-h}^{h}f(y,t)dy dt$. Note that the temporal average of $\partial \bar E_k/\partial t$ is zero due to the $2\pi$-periodicity of the shape function. In the figures to be discussed, we will plot the normalized energy terms, as shown in the second equation of \eqref{eq4d3}. The equation also serves as a posterior check for our calculation, that is, the sum of all the normalized terms is equal to $2\omega_i$ from the Floquet analysis.

\begin{figure}
	\centering
	\includegraphics[width=0.485\textwidth,trim= 0 0 -20 0,clip]{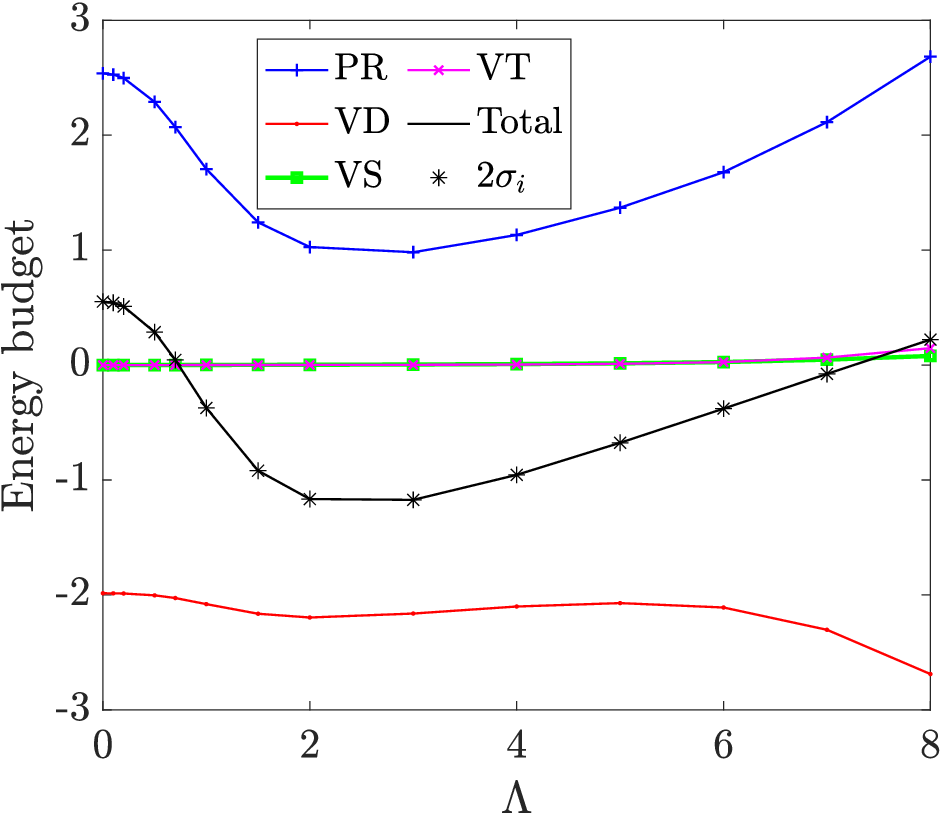}\put(-249,190){$(a)$}
	\includegraphics[width=0.48\textwidth,trim= 0 0 -20 0,clip]{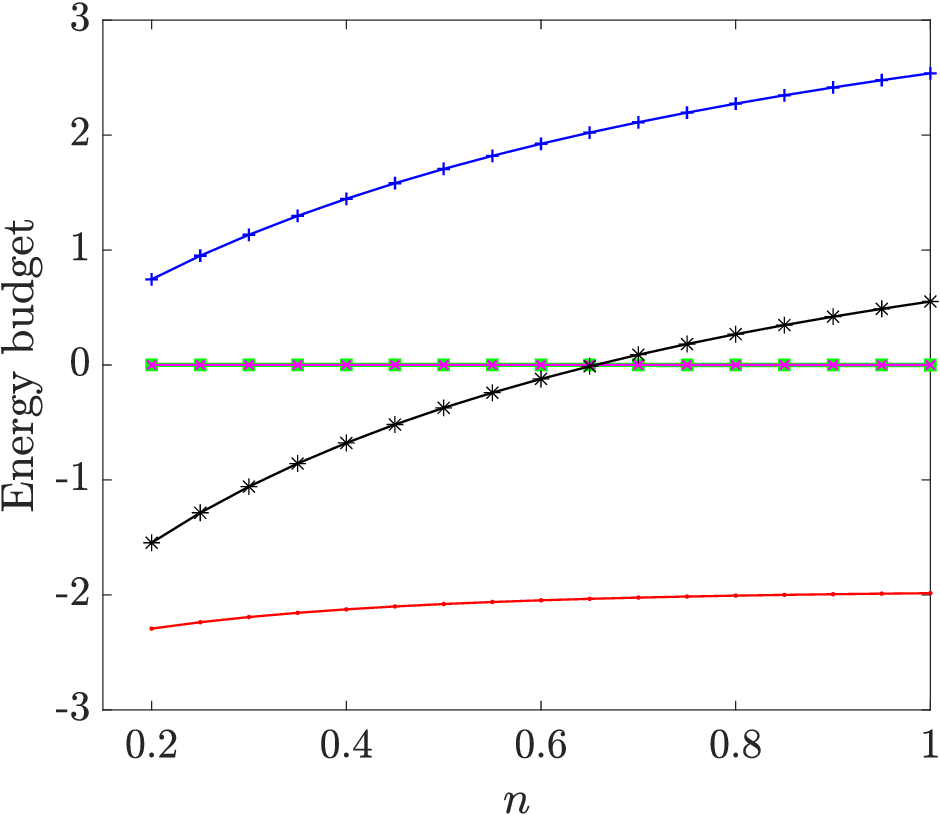}\put(-247,190){$(b)$}
	\caption{\color{black}Energy budget of the leading mode of the Stokes layer flow of Carreau fluids at $Re=700$: $(a)$ variation of the different terms with the Carreau number $\Lambda$ at $n=0.5$; $(b)$ variation with the power-law index $n$ at $\Lambda=1$. Note that at each parametric point, the linear growth rate is maximized over all wavenumber; the terms $\bar{\bar{f}}_{PR}, \bar{\bar{f}}_{VD}, \bar{\bar{f}}_{VS}, \bar{\bar{f}}_{VT}$ in this plot have been normalized by $\bar{\bar{E}}_{k}$ as in Eq. \eqref{eq4d3}. }
	\label{Fig:Energy_budget_lambda}
\end{figure}

Figure \ref{Fig:Energy_budget_lambda}$(a)$ presents the variations of the normalized terms with $\Lambda$ for the leading modes in the flows at $n=0.5$ and $Re=700$, together with the doubled linear growth rate discussed in figure \ref{Fig:GR_FR_varying_lambda}. From figure \ref{Fig:Energy_budget_lambda}$(a)$ one can see that the total summation of all the terms are exactly equal to $2\omega_i$, validating the inherent consistency of our energy analysis. Among these components, the energy production term exhibits the largest variation, closely following the variation of the linear growth rate, while the viscous dissipation term barely changes in the investigated $\Lambda$ range. The remaining terms are basically zero. This suggests that the stability/instability of the flow is dominated by the energy production mechanism, which is also the case regarding the effect of $n$ at fixed $\Lambda$ as shown in figure \ref{Fig:Energy_budget_lambda}$(b)$. 

\begin{figure}
	\centering
	\includegraphics[width=0.99\textwidth,trim= 55 25 40 40,clip]{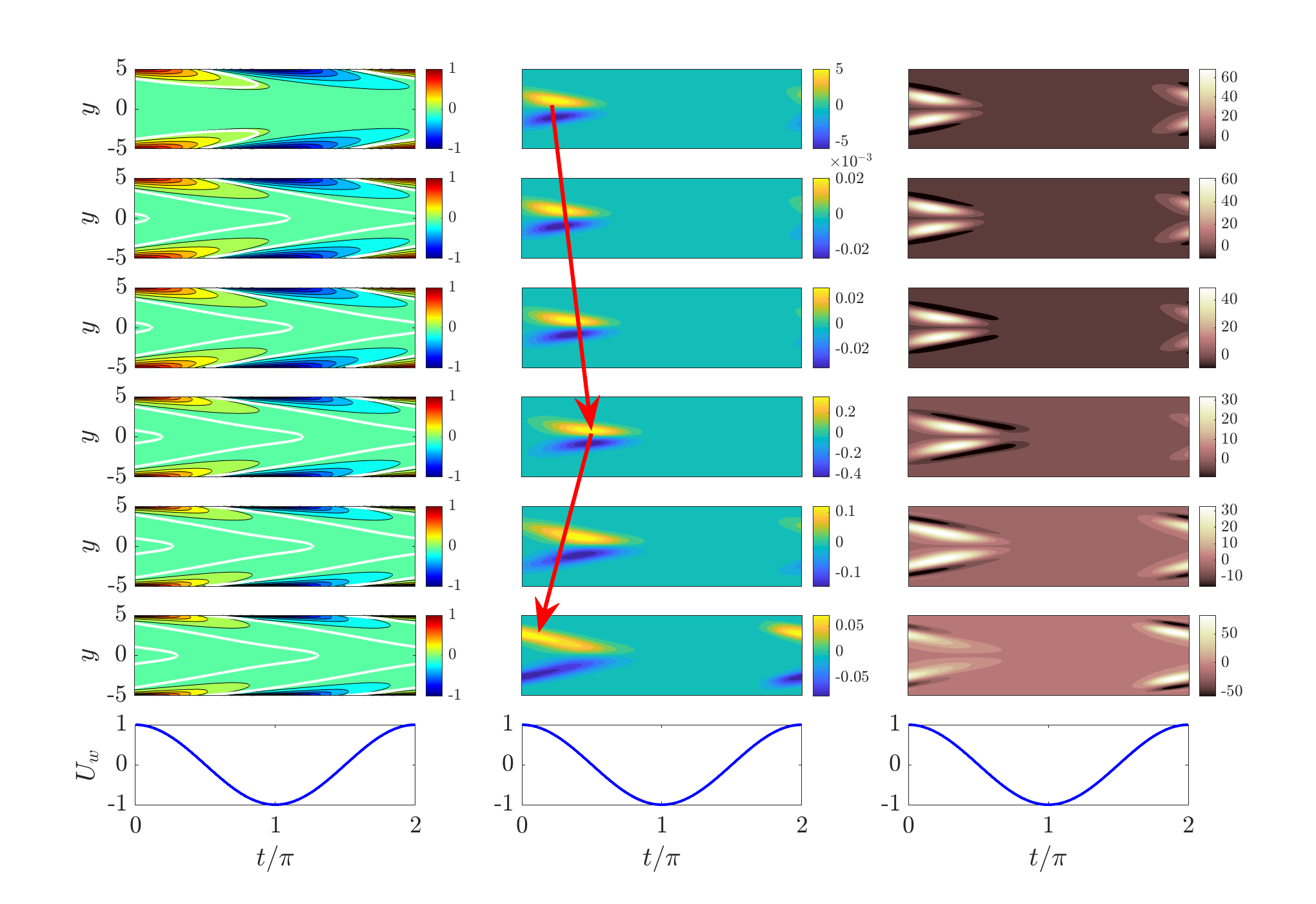}
	\caption{Mode patterns of the base flow velocity ($U_b$ in the first column), the real part of the multiplied perturbation field (Real($\tilde{u}^* \tilde{v}$) in the second column) and the resulting energy production ($f_{PR}$ in the third column) for the Stokes layer of Carreau fluids at $n=0.5$, $Re=700$ and varying $\Lambda$; {\color{black} for each $\Lambda$, the linear growth rate is maximized over all wavenumber $\alpha$. From top to bottom, $\Lambda=0,0.5,1,3,6,8$, respectively}. The last row shows the phase variation of the wall oscillation $\cos(t)$. Along the white curves in the first column plots the velocity is zero. Arrows in the second column demonstrate the phase change with increasing $\Lambda$. Note that at each $\Lambda$, we normalize the eigenfunction based on the value of its zeroth harmonic at the channel center ${v}^{(0)}(0)$ such that ${v}^{(0)}(0)$ has a zero phase angle and unity amplitude. }
	\label{Fig:phase_difference_varying_lambda}
\end{figure}

As the results in Figure~\ref{Fig:Energy_budget_lambda} highlight the dominant role of the energy production term, we now turn our attention to understanding why the production term $-\frac{Re}{2} \tilde{u}^* \tilde{v} \frac{\partial U_b}{\partial y}$ varies non-monotonically with \( \Lambda \). Drawing inspiration from the classical energy analysis of Newtonian channel flows, it is well known that the phase difference between the velocity perturbations \( u \) and \( v \) is responsible for the modal instability therein \cite{Schmid2001}. In our time-periodic base flow, however, not only do the perturbations \( u \) and \( v \) exhibit a phase difference, but the base flow $U_b$ or \( \partial U_b / \partial y \) itself evolves in time with its own phase. This necessitates a different perspective: instead of examining the phase between \( u \) and \( v \), we focus on the phase difference between the perturbation product \( \tilde{u}^{*} \tilde{v} \) and the time-varying base shear \( \partial U_b / \partial y \).

Figure \ref{Fig:phase_difference_varying_lambda} illustrates the base flow field, the real part of the multiplied perturbation field and the energy production distribution in the $y-t$ plane \dongdong{ for various values of \( \Lambda \) from 0 to 8}. We focus on the middle column in Figure \ref{Fig:phase_difference_varying_lambda}, as it shows most clearly the phase difference between the perturbation field and the time-periodic base flow; {\color{black} note that the six panels in the middle column correspond to unevenly set Carreau numbers of $\Lambda=0,0.5,1,3,6,8$, respectively}. The final row is included as a reference for the base flow phase evolution. The stabilizing effect observed when \( \Lambda \) increases from 0 to approximately 3 (see figure \ref{Fig:Energy_budget_lambda}) arises from an increasing phase mismatch between the disturbance and the base flow. This mismatch leads to a decreased overall energy production (see the increasing dark-colour area where $f_{PR}$ is negative in right column), thereby stabilizing the flow. {\color{black} At about $\Lambda=3$, the strongest perturbation, as indicated by the peak values of the Reynolds-stress-like term Real$(\tilde u^* \tilde v)$, happens at about $t=\pi/2$ which coincides with the moment when the wall velocity $U_w$ is approximately zero. Conversely, as \( \Lambda \) increases further from about 3 to 8, the phase alignment is gradually restored and the perturbation even lasts a longer time due to the strong shear-thinning effect at high $\Lambda$, see the middle column and note that the Carreau number $\Lambda$ is defined as the ratio of the viscosity relaxation time to the wall-oscillation penetration time in Eq. \eqref{eq:Lambda_def}. } This phase synchronisation enhances energy production (again see the colorbar in right panel), accounting for the destabilizing effect observed in strongly shear-thinning fluids. Such interpretation of the energy analysis seems not to be reported in the literature on the time-periodic flows.

Lastly, we discuss the implications of the newly-found instability. Both stabilizing and destabilizing effects of shear-thinning have been reported in the literature. For example, Nouar et al. \cite{Nouar2007Delaying} demonstrated that the shear-thinning stabilizes the flow especially when the tangent viscosity at the walls is used to define the Reynolds number. Delay of flow transition due to shear-thinning effects has also been observed in \cite{Kelly2020,Lacassagne2021}. On the other hand, Ashrafi \& Khayat \cite{Ashrafi2000} studied the chaos in the shear-thinned Taylor-Couette flow and Esmael et al. \cite{Esmael2010} reported a weakly turbulent state caused by the shear-thinning in a pipe, indicating the destabilizing effect of the shear-thinning in closed and shear flow systems, respectively. Our present calculations indicate a novel oscillatory flow instability in the weakly and strongly shear-thinning regimes in time-dependent flow systems, which complements the above studies on the effects of shear-thinning. Concerning practical applications, the discovery that shear-thinning can have a destabilizing effect on the oscillatory flow opens new possibilities for enhancing mixing, particularly in the strongly shear-thinning regime. For instance, in organ-on-chip platforms \cite{Low2021}, pulsatile flow is employed to mimic natural blood circulation in  human intestine, liver, skin and kidney equivalents \cite{Maschmeyer2015}. Making the fluid strongly shear-thinning may promote flow instabilities or even transition to turbulence, which is known to enhance mixing efficiency \cite{Stone2004,Niederkorn1994}, as recently demonstrated in \cite{Kumar2024}. Similarly, in the cosmetics and paint industries, shear-thinning fluids (such as lotions, gels, and latex) are commonly mixed using oscillatory agitation to avoid air entrainment while improving uniformity. The shear-thinning-induced instabilities identified in this work could therefore have practical implications for optimizing these processes.

\section{ Conclusion} \label{Conclusions}

This study conducted a Floquet analysis to investigate the stability of the finite Stokes layer flow in a plane channel filled with shear-thinning fluids, modeled using the Carreau rheological law. In studying the linear instability of this time-periodic flow, the first difficulty we encountered is the unavailability of the analytical solution to the laminar shear-thinning Stokes layer. To overcome this, we have conducted a binomial expansion of the Carreau model to arrive at an analytical expression of the base flow valid for small $\Lambda$, even though high-order solutions are still solved numerically. The good comparison of this method with the reference numerical solution validates the results of both methods. The Floquet analyzes based on the two base flows also present a good match at small $\Lambda$, indicating the usefulness of the obtained base flow using an expansion method, as it is much handier than the time-consuming numerical method. 

The parametric study revealed a non-monotonic effect of $\Lambda$ and a monotonic effect of $n$ on the shear-thinning oscillatory flow instability. Through an energy budget analysis, the underlying instability mechanism is revealed to be driven by phase matching between the base flow and the perturbation field, which enables the disturbance to extract energy from the base flow. This can occur when the shear-thinning is either weak or strong. In contrast, for moderate shear-thinning (intermediate $\Lambda$), a phase shift between the disturbance and the base flow leads to reduced energy production, stabilizing the flow. We have also explored the instability in an experimentally feasible manner, finding that solely changing the channel oscillation frequency exhibits a non-monotonic effect on the linear growth rate of disturbance as well. 

For future directions, the instability of oscillatory flows in viscoelastic fluids can be further explored. Since the Carreau model does not capture elastic effects, more sophisticated viscoelastic rheological models should be adopted. The present work isolates and examines the influence of shear-thinning alone, and therefore provides a useful reference point for future studies that incorporate elasticity.\\

\begin{acknowledgments}

This work is supported by Ministry of Education, Singapore via the grant WBS no.A-800117200-00. H.T. acknowledges the financial support from the National Natural Science Foundation of China (No. 12472271).

\end{acknowledgments}

\appendix

\section{Validation for the Poiseuille flow of Carreau fluids}\label{validationappendix}

\begin{figure}
	\centering
	\includegraphics[width=0.99\textwidth,trim= 0 0 0 0,clip]{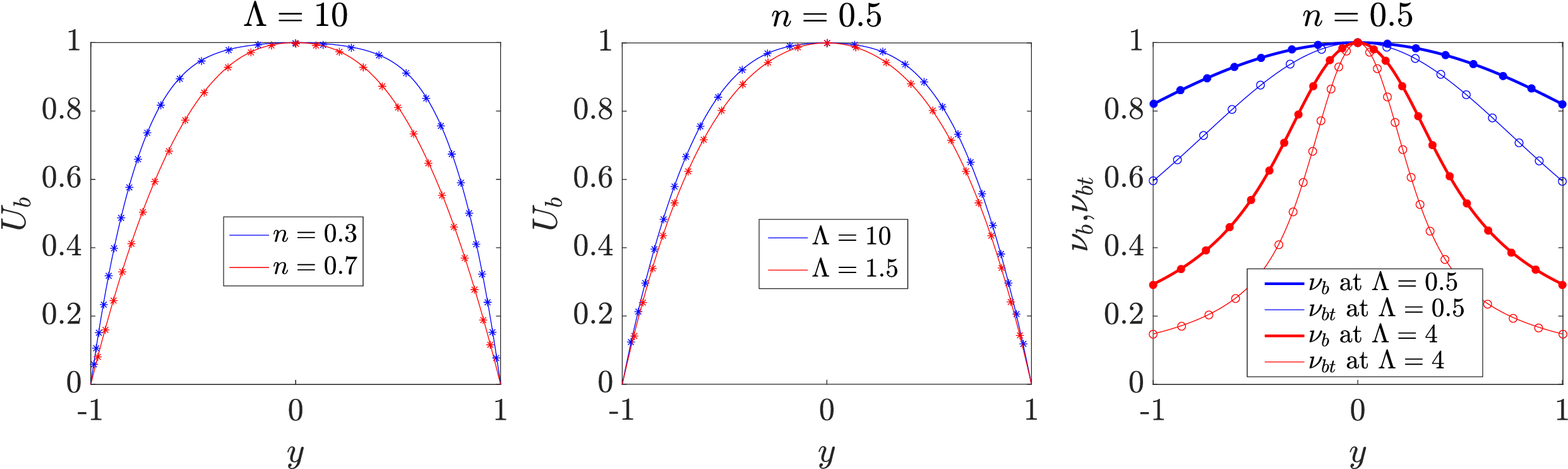} 
	\caption{Comparison of the base flow velocity and viscosity profiles calculated using the present code adjusted to the plane Poiseuille flow (solid lines) and that in \cite{Nouar2007Delaying} (symbols). }
	\label{Fig:compare_Nouar}
\end{figure}

To validate our code for calculating the numerical solution of the laminar base flow, we modified it to compute the velocity profiles in plane Poiseuille flow of Carreau fluids and compared the results with those in \cite{Nouar2007Delaying}. As shown in Figure \ref{Fig:compare_Nouar}, excellent agreement is achieved for various Carreau numbers $\Lambda$ and power-law indices $n$.

\dongdong{
\section{Numerical convergence tests}\label{convergence_appendix}

Table \ref{Tab:convergence1} shows an example of our numerical convergence test of the eigenvalue computations for weakly shear-thinning Carreau fluids.  Convergence to about 5 decimals of the most unstable eigenvalue is achieved for $N_f\gtrsim220$ at the parameter setting of $n=0.9$, $\Lambda=0.1$, $Re=700$ and $\alpha=0.41$. Thomas et al. \cite{Thomas2011} proposed that $N_f$ should be approximately greater than $0.8\alpha Re$ to guarantee numerical convergence for Newtonian Stokes layer flows. This corresponds to $0.8\alpha Re\approx 230$ for the present test, demonstrating that the their empirical estimation is still valid for weakly shear-thinning fluids. With enhancing the shear-thinning effects further to the parameter setting of  $n=0.5$ and $\Lambda=1$ but keeping $Re$ and $\alpha$ unchanged, $N_f\gtrsim220$ is still sufficient to obtain converged results; see table  \ref{Tab:convergence2}. {\color{black} For more strongly shear-thinning fluids at small $n$ and/or large $\Lambda$, such as the case shown in table \ref{Tab:convergence3} for $\Lambda=8$, $n=0.5$ and $\alpha=0.7$, approximately $N_f=390$ is required for convergence, which still corresponds to $0.8\alpha Re\approx 392$. Note that the largest linear growth rate is obtained at this wavenumber of $\alpha=0.7$; see FIG. \ref{Fig:GR_FR_varying_lambda}$(a)$.

These limited convergence tests suggest that the resolution requirement $N_f\gtrsim 0.8\alpha Re$ for convergence is likely to be also applicable for Stokes layer of shear-thinning fluids. Nevertheless, }
we also note that convergence is more difficult to be achieved with respect to $N_y$ when the shear-thinning effects become stronger, especially in terms of the mode frequency (real part of the eigenvalue). In the present study, we mainly focus on the mode growth rate and always choose sufficiently large $N_f$ (ranging from 300 to 390) and $N_y$ (ranging from 99 to 149) to guarantee its convergence at various parameters.
}

\begin{table}
	\begin{ruledtabular}  {\color{black}
			\begin{tabular}{p{1.5cm}p{3.5cm}p{3.5cm}p{3.5cm}p{3.5cm}} 
		     &  $N_y=39$  & $N_y=59$  & $N_y=79$ & $N_y=99$ \\ \hline
		$N_f=160$ &   $0.38912683 + 0.67126622$i  & $0.38379059 + 0.68002850$i & $0.38384657 + 0.68003154$i & $0.38387179 + 0.68003283$i \\
		$N_f=180$   &     $0.57049612+0.42028556$i  & $0.57026606+0.42812937$i & $0.57020336+0.42812966$i & $0.57017448 + 0.42813018$i \\
		$N_f=200$ &         $0.31342941+0.26993784$i & $0.31445416+0.27505530$i & $0.31438931+0.27505395$i & $0.31435963 + 0.27505343$i \\
		$N_f=220$ &         $0.31343272+0.26993006$i & $0.31445802+0.27504882$i & $0.31439317+0.27504747$i & $0.31436349 + 0.27504695$i \\
		$N_f=240$ &   $0.31343276+0.26993003$i & $0.31445802+0.27504882$i & $0.31439317+0.27504747$i & $0.31436352 + 0.27504700$i \\
		$N_f=260$ &  $0.31343272+0.26993006$i & $0.31445802+0.27504882$i &$0.31439312+0.27504758$i & $0.31436349 + 0.27504695$i \\ 
			\end{tabular} }
		\caption{\color{black}Numerical convergence of the most unstable eigenvalue for the Stokes layer of Carreau fluids at $n = 0.9$, $\Lambda=0.1$, $Re = 700$ and $\alpha = 0.41$.} 
		\label{Tab:convergence1}
	\end{ruledtabular}
\end{table}

\begin{table}
	\begin{ruledtabular} {\color{black}
			\begin{tabular}{p{1.5cm}p{3.5cm}p{3.5cm}p{3.5cm}p{3.5cm}} 
		     &  $N_y=79$  & $N_y=99$  & $N_y=119$ & $N_y=139$ \\ \hline
		$N_f=180$ &  $0.085102+0.313332$i & $0.089645+0.313556$i &$0.092050+0.313670$i & $0.093473+0.313735$i \\
		$N_f=200$ &   $0.092948-0.180231$i  & $0.098644-0.180203$i & $0.101662-0.180202$i & $0.103467-0.180192$i \\
		$N_f=220$   &     $0.085562-0.187291$i  & $0.091230-0.187315$i & $0.094238-0.187336$i & $0.096025-0.187352$i \\
		$N_f=240$ &         $0.085566-0.187289$i & $0.091230-0.187315$i & $0.094241-0.187334$i & $0.096024-0.187354$i \\
		$N_f=260$ &         $0.085563-0.187291$i & $0.091228-0.187316$i & $0.094234-0.187339$i & $0.096025-0.187352$i \\
		$N_f=280$ &   $0.085563-0.187291$i & $0.091229-0.187316$i & $0.094237-0.187339$i & $0.096025-0.187352$i \\ 
			\end{tabular} }
		\caption{\color{black}Numerical convergence of the most unstable eigenvalue for the Stokes layer of Carreau fluids at $n = 0.5$, $\Lambda=1$, $Re = 700$ and $\alpha = 0.41$.} 
		\label{Tab:convergence2}
	\end{ruledtabular}
\end{table}

\begin{table}
	\begin{ruledtabular} {\color{black}
			\begin{tabular}{p{1.5cm}p{3.5cm}p{3.5cm}p{3.5cm}p{3.5cm}}
		     &  $N_y=109$  & $N_y=129$  & $N_y=149$ & $N_y=169$ \\ \hline
		$N_f=350$ &  $0.478506+0.318554$i & $0.483081+0.319530$i &$0.485366+0.319968$i & $0.486923+0.320339$i \\
		$N_f=370$ &  $0.076052+0.114266$i & $0.080016+0.115261$i & $0.081999+0.115873$i & $0.081488+0.115976$i \\
		$N_f=390$ &  $0.076960+0.108972$i & $0.081355+0.109984$i & $0.084869+0.110466$i & $0.086027+0.110612$i \\
		$N_f=410$ &  $0.077090+0.108951$i & $0.082318+0.109993$i & $0.084513+0.110521$i & $0.085509+0.110655$i \\
			\end{tabular} }
		\caption{\color{black}Numerical convergence of the most unstable eigenvalue for the Stokes layer of Carreau fluids at $n = 0.5$, $\Lambda=8$, $Re = 700$ and $\alpha = 0.7$.} 
		\label{Tab:convergence3}
	\end{ruledtabular}
\end{table}

\section{ODEs and solutions for the base flow}\label{U11U13}

\subsection{Solutions at the order of $\Lambda^2$}

\subsubsection{Solving for $ U_{13}$}
We solve for the general solution of $U_{13}$ first, which is 
\begin{align}
6i U_{13}   &=  \frac{\partial^2    U_{13} }{\partial y^2}.
\end{align}
Assuming $U_{13}(y)=Ce^{my}$, we have
\begin{align}
6i Ce^{my}   =   m^2 Ce^{my} \ \ \ \ \  \text{   or  }\ \ \ \ \ m^2 -6i  = 0.
\end{align}
So $m = \pm \sqrt{6i}$ and the general solution $U_{13}^h$ reads
\begin{align}
U_{13}^h(y) = C_1 \cosh{\sqrt{6i} y}. 
\end{align}
For the particular solution $U_{13}^p$, we have the equation 
\begin{align}
6i U_{13}^{p}  -  \frac{\partial^2    U_{13}^{p} }{\partial y^2}  &=   \frac{n-1}{2} 6i(\frac{\partial  U_0 }{\partial y})^2  U_0  . 
\end{align}
On the other hand, $U_0 = \frac{1}{2}\frac{\cosh \sqrt{2i}y }{\cosh \sqrt{2i}h  }   $ and $\frac{\partial U_0}{\partial y} = \frac{\sqrt{2i}}{2}\frac{\sinh \sqrt{2i}y }{\cosh \sqrt{2i}h  }  $. By substitution, we have
\begin{align}
6i U_{13}^{p}  -  \frac{\partial^2    U_{13}^{p} }{\partial {y}^2}  &=  -K (2\cosh{3\sqrt{2i}y} -2\cosh{\sqrt{2i}y}   ),  \\
K & = \frac{n-1}{2} 12  (\frac{1 }{\cosh \sqrt{2i}h  })^2    \frac{1 }{8\cosh \sqrt{2i}h  } \frac{1}{8}.
\end{align}
Assume $U_{13}^p = A\cosh{3\sqrt{2i}y} +B \cosh{\sqrt{2i}y}  $, we can get $A=\frac{K}{6i }, B = \frac{K}{2i }$. Combining the general and particular solutions, we have
\begin{align}
U_{13} = U_{13}^h + U_{13}^p & =    C_1 \cosh{\sqrt{6i} y}   +  \frac{K}{6i }\cosh{3\sqrt{2i}y} +\frac{K}{2i } \cosh{\sqrt{2i}y}  
\end{align}
Imposing the homogeneous boundary conditions for $U_{13}$, i.e., $U_{13}(\pm h)=0$, we have
\begin{align}
 C_1 \cosh{\sqrt{6i} h}  +  \frac{K}{6i }\cosh{3\sqrt{2i}h} +\frac{K}{2i } \cosh{\sqrt{2i}h}  =0. 
\end{align}
So $C_1=\frac{iK}{6 } \frac{\cosh{3\sqrt{2i}h} +3 \cosh{\sqrt{2i}h}  }{  \cosh{\sqrt{6i}h}   } $. Finally, the solution is 
\begin{align}
U_{13} & =   \frac{iK}{6 } \frac{\cosh{3\sqrt{2i}h} +3 \cosh{\sqrt{2i}h}  }{  \cosh{\sqrt{6i}h}   }   \cosh{\sqrt{6i} y}  -  \frac{iK}{6 }(\cosh{3\sqrt{2i}y} +3 \cosh{\sqrt{2i}y}  ),\nonumber \\
\text{or} \ \ \ \ U_{13} & = \frac{n-1}{2} \frac{i}{8} \Big[  \frac{ \cosh{\sqrt{6i} y}}{ \cosh{\sqrt{6i}h}}  -  \frac{\cosh^3(\sqrt{2i}y  ) }{  \cosh^3(\sqrt{2i}h)   }\Big ]. 
\end{align}

\subsubsection{Solving for $U_{11}$}
We first solve for the general solution of $U_{11}$, which is governed by 
\begin{align}
2i U_{11}   &=  \frac{\partial^2    U_{11} }{\partial y^2}.  
\end{align}
Assuming $U_{11}(y)=Ce^{my}$, we have
\begin{align}
2i Ce^{my}   &=   m^2 Ce^{my} \ \ \ \ {or} \ \ \ \ m^2 -2i   = 0.
\end{align}
So $m = \pm \sqrt{2i}$ and the homogeneous solution $U_{11}^h$ reads
\begin{align}
U_{11}^h(y) = C_1 \cosh{\sqrt{2i} y}.  
\end{align}
Next, the particular solution is governed by
\begin{align}
2i U_{11}^{p}  -  \frac{\partial^2    U_{11}^{p} }{\partial y^2}  &=   \frac{n-1}{2} 12i \frac{\partial  U_0 }{\partial y} \frac{\partial  U_0^* }{\partial y} U_0    - \frac{n-1}{2} 6i (\frac{\partial  U_0 }{\partial y} )^2 U_0^*.
\end{align}
By substituting the expression of $U_0$, we can obtain
\begin{align}
2i U_{11}^{p}  -  \frac{\partial^2    U_{11}^{p} }{\partial {y}^2}  =  8K'  \Big[ 2i  \sinh \sqrt{2i}y & \sinh \sqrt{2/i}y \cosh \sqrt{2i}y  +  \sinh \sqrt{2i}y \sinh \sqrt{2i}y \cosh \sqrt{2/i}y  \Big],  \nonumber \\
\text{where} \ \ \ K'&=  \frac{n-1}{2} 12(\frac{1 }{\cosh \sqrt{2i}h  })^2    \frac{1 }{8\cosh \sqrt{2/i}h  } \frac{1}{8},  \\
2i U_{11}^{p}  -  \frac{\partial^2    U_{11}^{p} }{\partial {y}^2}  =  K'  \Big[ 2(1+2i)& \cosh(3+i)y - 4\cosh(1-i)y +2 (1-2i)\cosh(1+3i)y \Big]. 
\end{align}
Assume $U_{11}^p = A\cosh(3+i)y +B \cosh(1-i)y  +C \cosh(1+3i)y  $, we can get $A = \frac{K'}{10}(-4-3i), \ \ \ B = iK', \ \ \ C = \frac{K'}{10}(4-3i)$. 
So, we have
\begin{align}
U_{11} = U_{11}^h + U_{11}^p  &=    C_1 \cosh{\sqrt{2i} y} +  \frac{K'}{10}(-4-3i)\cosh(3+i)y \nonumber \\
 &+iK' \cosh(1-i)y  +\frac{K'}{10}(4-3i) \cosh(1+3i)y.    
\end{align}
Imposing the homogeneous boundary condition for $U_{11}$, i.e., $U_{11}(\pm h)=0$, we have
\begin{align}
C_1 \cosh{\sqrt{2i} h}  +  \frac{K'}{10}(-4-3i)\cosh(3+i)h +iK' \cosh(1-i)h  +\frac{K'}{10}(4-3i) \cosh(1+3i)h  =0.  \nonumber 
\end{align}
So $C_1 = \frac{  \frac{K'}{10}(4+3i)\cosh(3+i)h -iK' \cosh(1-i)h  -\frac{K'}{10}(4-3i) \cosh(1+3i)h   }{\cosh(\sqrt{2i}h)}$. By setting $K = \frac{K'}{\cosh\sqrt{2i}h}$, we can obtain the final solution as
\begin{align}
U_{11}(y) & =   \frac{K}{10}(4+3i)\Big(\cosh(3+i)h\cosh{\sqrt{2i} y} - \cosh(3+i)y\cosh \sqrt{2i}h \Big)  \nonumber \\
 & -iK \Big( \cosh(1-i)h \cosh{\sqrt{2i} y} - \cosh(1-i)y\cosh \sqrt{2i}h \Big) \nonumber \\
 & - \frac{K}{10}(4-3i)\Big( \cosh(1+3i)h   \cosh{\sqrt{2i} y} -  \cosh(1+3i)y\cosh \sqrt{2i}h \Big)    \\
\text{with} \ \ \ K &= \frac{n-1}{2} \frac{3}{16}(\frac{1 }{\cosh \sqrt{2i}h  })^3    \frac{1 }{\cosh \sqrt{2/i}h  }.  
\end{align}

\subsection{ODEs for the solutions at the order of $\Lambda^4$}\label{app:order4}

The corresponding ODEs for $U_{21}$, $U_{23}$ and $U_{25}$ are presented as follows.
We can get for $U_{21}$
\begin{align}
i U_{21}  = \frac{1}{2}\frac{\partial }{\partial y}[ \frac{\partial  U_{21} }{\partial y}+ \frac{n-1}{2} 3(\frac{\partial   U_{0} }{\partial y}^2\frac{\partial   U^*_{11} }{\partial y} +\frac{\partial   U^*_{0} }{\partial y}^2 \frac{\partial   U_{13} }{\partial y}  + 2\frac{\partial   U_{0} }{\partial y}\frac{\partial   U^*_{0} }{\partial y} \frac{\partial   U_{11} }{\partial y}  )
+  \frac{(n-1)(n-3)}{8} 10 (\frac{\partial U_0 }{\partial y})^3(\frac{\partial U^*_0 }{\partial y})^2 ].
\end{align}

We can get for $U_{23}$
\begin{align}
3i U_{23}  = \frac{1}{2}\frac{\partial }{\partial y}\Big[ \frac{\partial  U_{23} }{\partial y} &+\frac{n-1}{2} 3(\frac{\partial   U_{0} }{\partial y}^2 \frac{\partial   U_{11} }{\partial y}   + 2\frac{\partial   U_{0} }{\partial y}\frac{\partial   U^*_{0} }{\partial y} \frac{\partial   U_{13} }{\partial y}  )
+   \frac{(n-1)(n-3)}{8} 5 (\frac{\partial U_0 }{\partial y})^4 \frac{\partial U^*_0 }{\partial y} \Big]. 
\end{align}

We can get for $U_{25}$
\begin{align}
5i U_{25}  = \frac{1}{2}\frac{\partial }{\partial y}\Big[ \frac{\partial  U_{25} }{\partial y}+ \frac{n-1}{2} 3(\frac{\partial   U_{0} }{\partial y}^2 \frac{\partial   U_{13} }{\partial y}  )  +   \frac{(n-1)(n-3)}{8}  (\frac{\partial U_0 }{\partial y})^5 \Big].
\end{align}

\subsection{At the order of $\Lambda^6$}\label{app:order6}
\dongdong{
At this order,  we have
\begin{align}\label{Lambda6Eq}
 \frac{\partial    U_{b3} }{\partial t}  = \frac{1}{2}\frac{\partial }{\partial y}\Big[ \frac{\partial U_{b3} }{\partial y}&+ 3\frac{n-1}{2} \Big[ \big(\frac{\partial  \Ubz}{\partial y} \big)^2\big(\frac{\partial  \Ubt}{\partial y} \big)  + \big(\frac{\partial  \Ubz}{\partial y} \big)\big(\frac{\partial  \Ubo}{\partial y} \big)^2     \Big] \nonumber \\
 &+  5 \frac{(n-1)(n-3)}{8} (\frac{\partial \Ubz }{\partial y})^4\big(\frac{\partial  \Ubo}{\partial y} \big) + \frac{(n-1)(n-3)(n-5)}{48} (\frac{\partial \Ubz }{\partial y})^7 \Big].
\end{align}
A procedure similar to  can be followed with $U_{b3}(y,t) = U_{31} e^{it}+ U_{33} e^{3it}+ U_{35}e^{5it}+ U_{37}e^{7it} + c.c.$ and the corresponding ODEs for $U_{31}$, $U_{33}$, $U_{35}$ and $U_{37}$ to be solved numerically are as follows.}

We can get for $U_{31}$
\begin{align}
i U_{31}  = \frac{1}{2}\frac{\partial }{\partial y} \Bigl\{ \frac{\partial  U_{31} }{\partial y}  +\frac{n-1}{2} 3\Big[    \big(\frac{\partial  U_0}{\partial y} \big)^2\big(\frac{\partial  U^*_{21}}{\partial y} \big) +  2\big(\frac{\partial  U_0}{\partial y} \big)\big(\frac{\partial  U^*_0}{\partial y} \big)\big(\frac{\partial  U_{21}}{\partial y} \big)  \nonumber \\
+ \big(\frac{\partial  U^*_0}{\partial y} \big)^2\big(\frac{\partial  U_{23}}{\partial y} \big) + \frac{\partial  U^*_0}{\partial y} \big(\frac{\partial  U_{11}}{\partial y} \big)^2    + 2    \frac{\partial  U_0}{\partial y} \frac{\partial  U_{13}}{\partial y} \frac{\partial  U^*_{13}}{\partial y} + 2    \frac{\partial  U_0}{\partial y} \frac{\partial  U_{11}}{\partial y} \frac{\partial  U^*_{11}}{\partial y} + 2    \frac{\partial  U^*_0}{\partial y} \frac{\partial  U_{13}}{\partial y} \frac{\partial  U^*_{11}}{\partial y}       \Big]  \nonumber \\
 + \frac{(n-1)(n-3)}{8} 5 \Big[  (\frac{\partial U_0 }{\partial y})^4\big(\frac{\partial  U^*_{13}}{\partial y} )  +4 (\frac{\partial U_0 }{\partial y})^3\frac{\partial U^*_0 }{\partial y}\big(\frac{\partial  U^*_{11}}{\partial y} )  +4 \frac{\partial U_0 }{\partial y} (\frac{\partial U^*_0 }{\partial y})^3\big(\frac{\partial  U_{13}}{\partial y} )  \nonumber  \\
 + 6(\frac{\partial U_0 }{\partial y})^2(\frac{\partial U^*_0 }{\partial y})^2 \big(\frac{\partial  U_{11}}{\partial y} )  \Big]
+   \frac{(n-1)(n-3)(n-5)}{48} 35 (\frac{\partial U_0 }{\partial y})^4 (\frac{\partial U^*_0 }{\partial y})^3 \Bigl\}.
\end{align}

We can get for $U_{33}$
\begin{align}
3i U_{33}  = \frac{1}{2}\frac{\partial }{\partial y}\Bigl\{ \frac{\partial  U_{33} }{\partial y} +\frac{n-1}{2} 3\Big[    \big(\frac{\partial  U_0}{\partial y} \big)^2\big(\frac{\partial  U_{21}}{\partial y} \big) +  2\big(\frac{\partial  U_0}{\partial y} \big)\big(\frac{\partial  U^*_0}{\partial y} \big)\big(\frac{\partial  U_{23}}{\partial y} \big) \nonumber \\
 + \big(\frac{\partial  U^*_0}{\partial y} \big)^2\big(\frac{\partial  U_{25}}{\partial y} \big)  + \frac{\partial  U_0}{\partial y} \big(\frac{\partial  U_{11}}{\partial y} \big)^2    + 2    \frac{\partial  U_0}{\partial y} \frac{\partial  U_{13}}{\partial y} \frac{\partial  U^*_{11}}{\partial y}+ 2    \frac{\partial  U^*_0}{\partial y} \frac{\partial  U_{13}}{\partial y} \frac{\partial  U_{11}}{\partial y}  \Big]  \nonumber \\
 + \frac{(n-1)(n-3)}{8} 5 \Big[  (\frac{\partial U_0 }{\partial y})^4\big(\frac{\partial  U^*_{11}}{\partial y} )  +4 (\frac{\partial U_0 }{\partial y})^3\frac{\partial U^*_0 }{\partial y}\big(\frac{\partial  U_{11}}{\partial y} )  + 6(\frac{\partial U_0 }{\partial y})^2(\frac{\partial U^*_0 }{\partial y})^2 \big(\frac{\partial  U_{13}}{\partial y} )  \Big] \nonumber\\
+  \frac{(n-1)(n-3)(n-5)}{48} 21 (\frac{\partial U_0 }{\partial y})^5 (\frac{\partial U^*_0 }{\partial y})^2 \Bigl\}. 
\end{align}

We can get for $U_{35}$
\begin{align}
5i U_{35}  = \frac{1}{2}\frac{\partial }{\partial y} \Bigl\{ \frac{\partial  U_{35} }{\partial y}  + \frac{n-1}{2} 3\Big[    \big(\frac{\partial  U_0}{\partial y} \big)^2\big(\frac{\partial  U_{23}}{\partial y} \big) +  2\big(\frac{\partial  U_0}{\partial y} \big)\big(\frac{\partial  U^*_0}{\partial y} \big)\big(\frac{\partial  U_{25}}{\partial y} \big)
 + \big(\frac{\partial  U^*_0}{\partial y} \big)\big(\frac{\partial  U_{13}}{\partial y} \big)^2    + 2\big(\frac{\partial  U_0}{\partial y} \big)  \frac{\partial  U_{13}}{\partial y}  \frac{\partial  U_{11}}{\partial y}       \Big]  \nonumber \\
 + \frac{(n-1)(n-3)}{8} 5 \Big[  (\frac{\partial U_0 }{\partial y})^4\big(\frac{\partial  U_{11}}{\partial y} ) + 4(\frac{\partial U_0 }{\partial y})^3\frac{\partial U^*_0 }{\partial y} \big(\frac{\partial  U_{13}}{\partial y} )  \Big] 
+  \frac{(n-1)(n-3)(n-5)}{48} 7 (\frac{\partial U_0 }{\partial y})^6 \frac{\partial U^*_0 }{\partial y}  \Bigl\}. 
\end{align}

We can get for $U_{37}$
\begin{align}
7i U_{37}  = \frac{1}{2}\frac{\partial }{\partial y} \Bigl\{  \frac{\partial  U_{37} }{\partial y} +\frac{n-1}{2} 3\Big[    \big(\frac{\partial  U_0}{\partial y} \big)^2\big(\frac{\partial  U_{25}}{\partial y} \big)  + \big(\frac{\partial  U_0}{\partial y} \big)\big(\frac{\partial  U_{13}}{\partial y} \big)^2         \Big]   &+ \frac{(n-1)(n-3)}{8} 5(\frac{\partial U_0 }{\partial y})^4\big(\frac{\partial  U_{13}}{\partial y} ) \nonumber \\
&+  \frac{(n-1)(n-3)(n-5)}{48}  (\frac{\partial U_0 }{\partial y})^7  \Bigl\}. 
\end{align}

%\nocite{*}

\bibliography{BibRef}% Produces the bibliography via BibTeX.

\end{document}